\documentclass[
    aps,
    prab,
    reprint,
    superscriptaddress,
    nofootinbib,
    groupedaddress
]{revtex4-2}

\usepackage{amsmath,amssymb}
\usepackage{graphicx}
\usepackage{booktabs}
\usepackage{siunitx}
\usepackage{bm}
\usepackage{microtype}
\usepackage{float}

\begin{document}


\title{Loss Factor of Axisymmetric and Rectangular Tapered Collimators from Optical to Inductive Regimes}


\author{Md Wahidur Rahman}
\email[Corresponding author: ]{mrahman5@bnl.gov}

\author{Alexei Blednykh}
\author{Gang Wang}
\author{Xiaofeng Gu}

\affiliation{
Brookhaven National Laboratory,
Electron-Ion Collider,
Upton, New York 11973-5000, USA
}


\begin{abstract}
In this work, we investigate the loss factor of axisymmetric
and rectangular tapered collimators over a broad range of
bunch lengths and geometrical parameters using the
electromagnetic simulation codes ECHO and GdfidL.
The calculations are performed for Gaussian bunches of
rms length $\sigma_s$, with the collimator geometry
characterized by the minimum and maximum apertures
$b$ and $d$, taper length $L$, and chamber width $W$
for rectangular structures.
For short bunches and tapered collimators with $d-b>b$, 
we extend the phenomenological formula
previously obtained for axisymmetric
to rectangular tapered collimators, describing the dependence of
the loss factor on bunch length and taper geometry through
an arctangent function of the scaling parameter
$\mu d^2/(\sigma_s L)$.
We demonstrate that this parameter is related to a Fresnel number
and, by combining diffraction theory with the
electrodynamics of the beam-induced image charge,
analytically explain the factor of two between the
coefficients $\mu=0.2$ and $\mu=0.4$ for axisymmetric
and rectangular collimators, respectively.
The resulting expression is applicable to rectangular
collimators with $W/(2d) \geq 1.8$, extending its validity
beyond the idealized flat geometry.
For longer bunches, we identify the transition from the
optical to the inductive regime and derive an analytical
expression for the loss factor based on the cutoff
wavenumber of the lowest relevant electromagnetic mode.
\end{abstract}

\maketitle

\section{Introduction}

\begin{figure}[H]
    \centering
    \includegraphics[width=1\columnwidth]{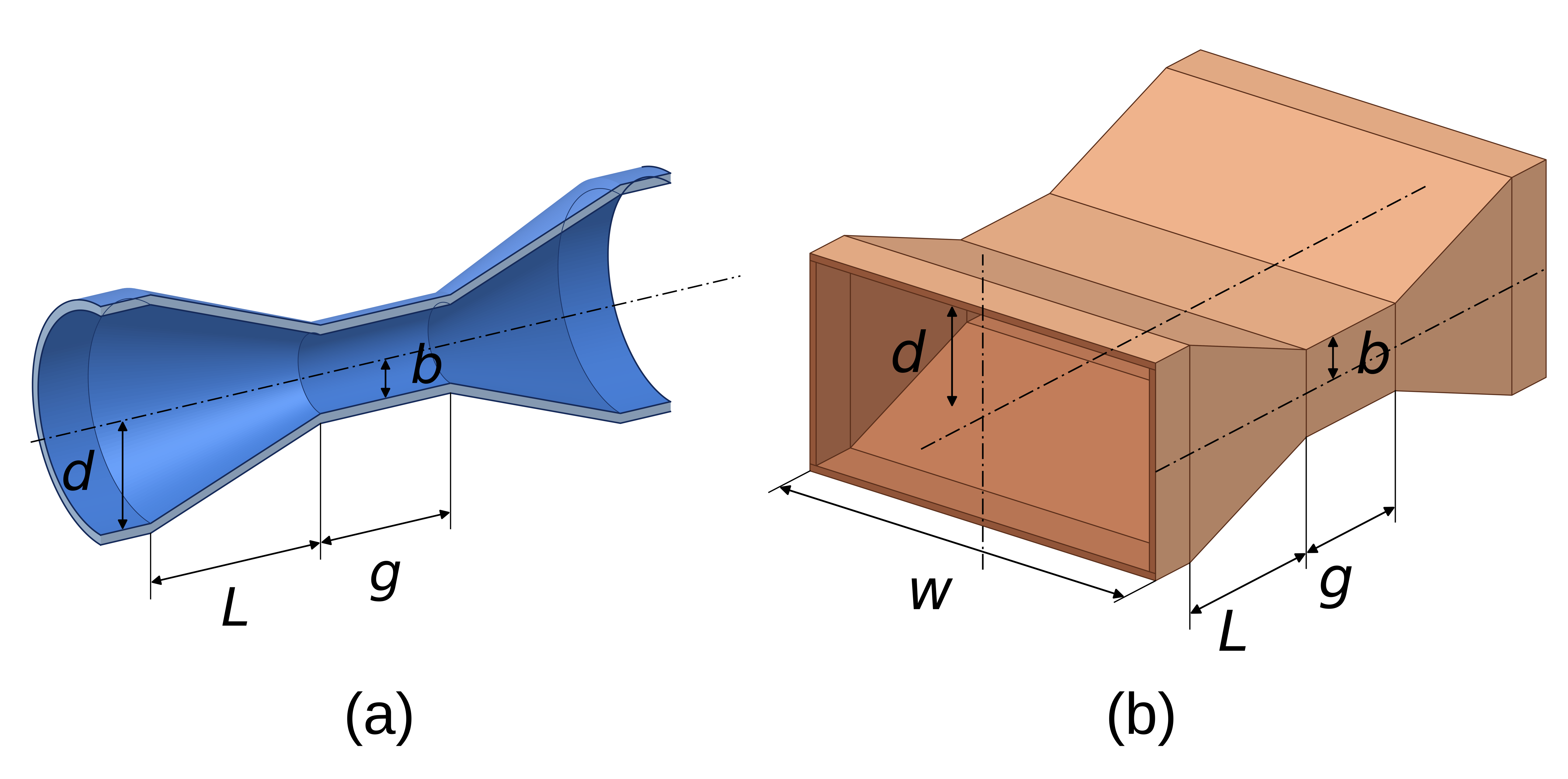}
   \caption{(a) Tapered axisymmetric collimator. (b) Tapered rectangular collimator.}
    \label{fig:geometry}
\end{figure}

The vacuum system of an accelerator contains numerous components with changes in cross section. Collimators are among the most critical of these components, as they intercept the beam halo, localize particle losses, and protect sensitive equipment and detectors. In a collider complex such as the Electron-Ion Collider (EIC), collimators in the Electron Storage Ring (ESR) protect the interaction region and reduce detector backgrounds~\cite{EICCDR2021}. To avoid abrupt changes in cross section at the narrow aperture and to reduce the beam coupling impedance, the collimator jaws are tapered over a length $L$ from the chamber half-aperture $d$ to the minimum half-aperture $b$ (see Fig.~\ref{fig:geometry}). A charged-particle bunch passing through such a transition excites electromagnetic wakefields that can lead to energy loss and beam quality degradation~\cite{Blednykh2010,Blednykh2011,Blednykh2021,Stupakov2007Optical,Palumbo1995}. The strength of these wakefields, which must be minimized, depends on the bunch length, transition regime, and collimator geometry, including the aperture dimensions, taper length, and cross-sectional shape~\cite{Chao1993,Zotter1998}.

The longitudinal impedance of tapered collimators is known analytically in two limits. At low frequencies it is inductive, as derived by Yokoya for round tapers~\cite{Yokoya1990} and by Stupakov for arbitrary cross sections~\cite{Stupakov2007}. At high frequencies, the optical approximation gives a real impedance determined by the transverse geometry alone~\cite{Stupakov2007Optical,Bane2007}. The conventional optical approximation is valid when the transition length is short compared with the catch-up distance at high frequency~\cite{Bane2007}. Between these limits, Stupakov, Bane and Zagorodnov~\cite{Stupakov2007Optical} formulated the problem with a parabolic (paraxial) equation, and Podobedov and Stupakov~\cite{StupakovPodobedov2010} computed the impedance of round collimators with an eigenmode method. Based on ECHO~\cite{Zagorodnov2005,Zagorodnov2006} and GdfidL~\cite{BrunsGdfidL} simulations, Blednykh and Krinsky~\cite{Blednykh2010,Blednykh2011} found that the loss factor of short bunches follows a phenomenological arctan law 
in $d^2/(\sigma_s L)$, with coefficients
$\mu_{\mathrm{rnd}} = 0.2$ and
$\mu_{\mathrm{flt}} = 0.4$ for round and
flat collimators, respectively.

For a Gaussian bunch with rms length $\sigma_s$, the loss factor is defined as
\begin{equation}
k_{\parallel}(\sigma_s)
=
\frac{\Delta E}{Q^2}
=
\int_{-\infty}^{\infty}
W_{\parallel}(s)\,\lambda(s)\,ds ,
\label{eq:longitudinal_loss_factor}
\end{equation}
where $\Delta E$ is the energy lost by the bunch, $Q$ is the total charge, $\lambda(s)$ is the normalized longitudinal charge distribution,
and $W_{\parallel}(s)$ is the longitudinal wake potential.

The tapered axisymmetric and rectangular collimator are shown in Fig.~1(a), and in Fig.~1(b), respectively. In the axisymmetric collimator, \(b\) and \(d\) denote the smaller and larger radii, respectively. In the rectangular collimator, \(b\) and \(d\) denote the smaller and larger vertical half-apertures, while \(W\) denotes the full chamber width. In both geometries, \(L\) is the taper length and \(g\) is the length of the narrow central section. 

In this paper, the loss factor of round and flat tapered collimators is computed with GdfidL and ECHO and described by a single formula, Eq.~\eqref{eq:two_regime_intro}, that covers the optical and inductive regimes. The resulting two-regime loss factor model is given by

\begin{equation}
k_{\parallel,\mathrm{tpr}}(\sigma_s)
=
\left\{
\begin{array}{@{}l@{\hspace{-0.0em}}l@{}}
\displaystyle
k_{\parallel,\mathrm{step}}(\sigma_s)
\left[
\frac{2}{\pi}
\tan^{-1}\!\left(
\frac{\mu d^2}{\sigma_s L}
\right)
\right]^2,
&
\begin{array}{l}
\sigma_s\leq\sigma_i\\
\text{(optical)}
\end{array}
\\[12pt]
\displaystyle
\begin{aligned}
&
k_{\parallel,\mathrm{step}}(\sigma_i)
\left[
\frac{2}{\pi}
\tan^{-1}\!\left(
\frac{\mu d^2}{\sigma_i L}
\right)
\right]^2
\\[-1pt]
&\times
\left(\frac{\sigma_i}{\sigma_s}\right)^2
\exp\!\left[
-k_c^2\left(\sigma_s^2-\sigma_i^2\right)
\right],
\end{aligned}
&
\begin{array}{l}
\sigma_s>\sigma_i\\
\text{(inductive)}
\end{array}
\end{array}
\right.
\label{eq:two_regime_intro}
\end{equation}
\\
where $k_{\parallel,\mathrm{step}}$ is the loss factor of step collimator, $\sigma_i=1/(\sqrt{2\ln 2}\,k_c)$ defines the optical to inductive regime transition bunch length, and $k_c$ is the cutoff wavenumber of the lowest relevant electromagnetic mode of the collimator, with $k_c^{\mathrm{rnd}}=j_{01}/d$ for the round and $k_c^{\mathrm{flat}}=\pi\sqrt{1/W^2+1/(2d)^2}$ for the flat geometry, respectively. Here, $j_{01}$ is the first zero of the Bessel function of the first kind, $J_0$. Thus, the cutoff wavenumber, $k_c$  depends primarily on $d$ for the round collimator, and on both $d$ and chamber width $W$ for the flat collimator.

The factor of two between these two collimators,
$\mu_{\mathrm{flt}}/\mu_{\mathrm{rnd}}\approx2$ discussed in Ref~\cite{Blednykh2011}, is derived here analytically from a Fresnel number~\cite{Siegman1986,FresnelDiffraction,PaschottaFresnel} defined by the transverse extent of the image charge, with $N_F^{\mathrm{rnd}}=d^2/(2\sigma_s L)$
and $N_F^{\mathrm{flt}}=d^2/(\sigma_s L)$.
In this interpretation, the bunch length provides the characteristic longitudinal scale, the taper length represents the propagation distance, and the transverse extent of the image charge distribution defines the effective aperture. The taper arguments can be
written as $\mu_{\mathrm{rnd}}d^2/(\sigma_sL)=0.4N_F^{\mathrm{rnd}}$
and
$\mu_{\mathrm{flt}}d^2/(\sigma_sL)=0.4N_F^{\mathrm{flt}}$.
Thus,
\begin{equation}
\frac{\mu_{\mathrm{flt}}}{\mu_{\mathrm{rnd}}}
=
\frac{N_F^{\mathrm{flt}}}{N_F^{\mathrm{rnd}}}
\approx
2,
\label{eq:mu_ratio}
\end{equation}
showing the empirical relation~\cite{Blednykh2011}
is consistent with the corresponding Fresnel number ratio. As the bunch length increases, the system departs from the
high frequency optical limit and diffraction effects become more
important~\cite{Stupakov2007Optical}.

\begin{figure}[H]
    \centering
    \includegraphics[width=\columnwidth]{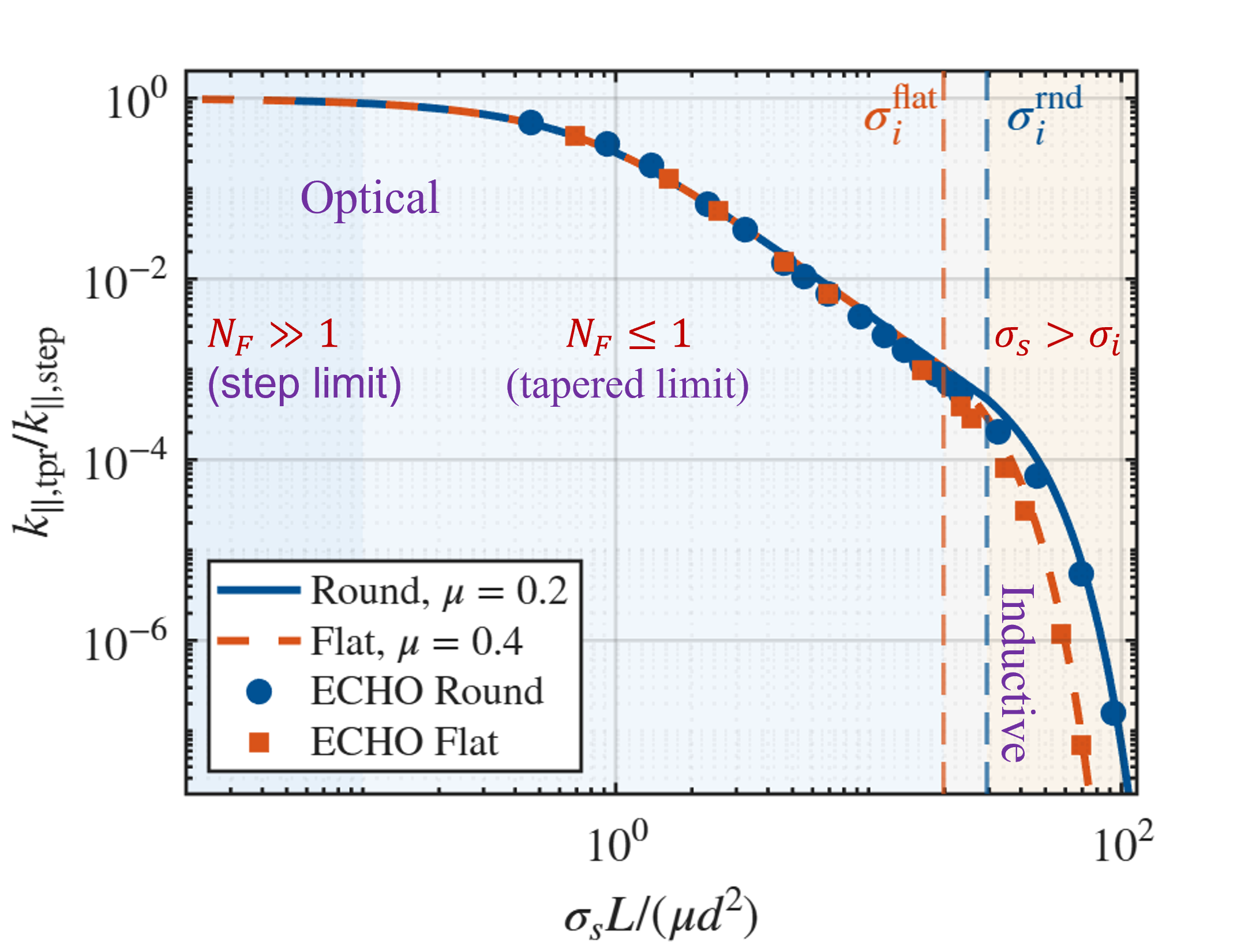}
    \caption{Normalized loss factor for round and flat tapered collimators obtained from the loss factor model (Eq.~\eqref{eq:two_regime_intro}) and numerical simulations on the scaled Fresnel axis. The reference geometry is $b=4.6~\mathrm{mm}$, $d=18~\mathrm{mm}$, $L=300~\mathrm{mm}$, and $W=67~\mathrm{mm}$. The vertical dashed lines indicate the optical and inductive transition
    $\sigma_i^{\mathrm{rnd}}=6.34$~mm and $\sigma_i^{\mathrm{flt}}=8.55$~mm. The inductive regime starts at $\sigma_s=\sigma_i$, where $N_F=N_F(\sigma_i)\propto d/L$ given by Eqs.~\eqref{eq:NF_round_transition} and~\eqref{eq:NF_flat_transition}. The shaded regions indicate the optical (light blue) regime, including the step and tapered limits, and the inductive (light brown) regime.}
    \label{fig:norm_mu}
\end{figure}

We compare the normalized loss factor for the round and flat tapered collimators using the Fresnel number scaling $\sigma_s L/(\mu d^2)$ shown in Fig.~\ref{fig:norm_mu}. For $N_F\gg1$, both geometries approach the abrupt-step limit. As $N_F$ decreases ($N_F\lesssim1$), the loss factors follow a similar tapered optical scaling. For sufficiently long bunches ($\sigma_s>\sigma_i$), the behavior transitions to the inductive regime, where the curves separate according to the geometry-dependent transition bunch lengths $\sigma_i^{\mathrm{rnd}}$ and $\sigma_i^{\mathrm{flt}}$.

In this work, we extend the loss factor analysis to a broad range of bunch lengths for both axisymmetric and rectangular tapered collimators. First, we investigate the dependence of the loss factor on bunch length and geometric parameters and describe the transition from the optical to the inductive regime. We then examine the relation between the conventional catch-up distance criterion and the short-bunch behavior. Next, we develop a two-regime loss factor model valid from short to long
bunches and compare it with numerical results from ECHO and GdfidL. Finally, we apply the model to estimate beam-induced power loss in an Electron Storage Ring (ESR) collimator of the Electron-Ion Collider (EIC).

\section{Fresnel Number of a Tapered Collimators}
\label{subsec:fresnel}

The phenomenological arctangent taper factor in Eq.~\eqref{eq:two_regime_intro} is derived analytically in this work through Fresnel-number analysis. In diffraction theory, the Fresnel number $N_F^{\mathrm{opt}}=a^2/(\lambda L)$
compares a transverse aperture $a$ with the diffraction spread of a wave
of wavelength $\lambda$ over a propagation distance
$L$~\cite{Siegman1986,FresnelDiffraction,PaschottaFresnel}.

\begin{figure}[H]
    \centering
    \includegraphics[width=\columnwidth]{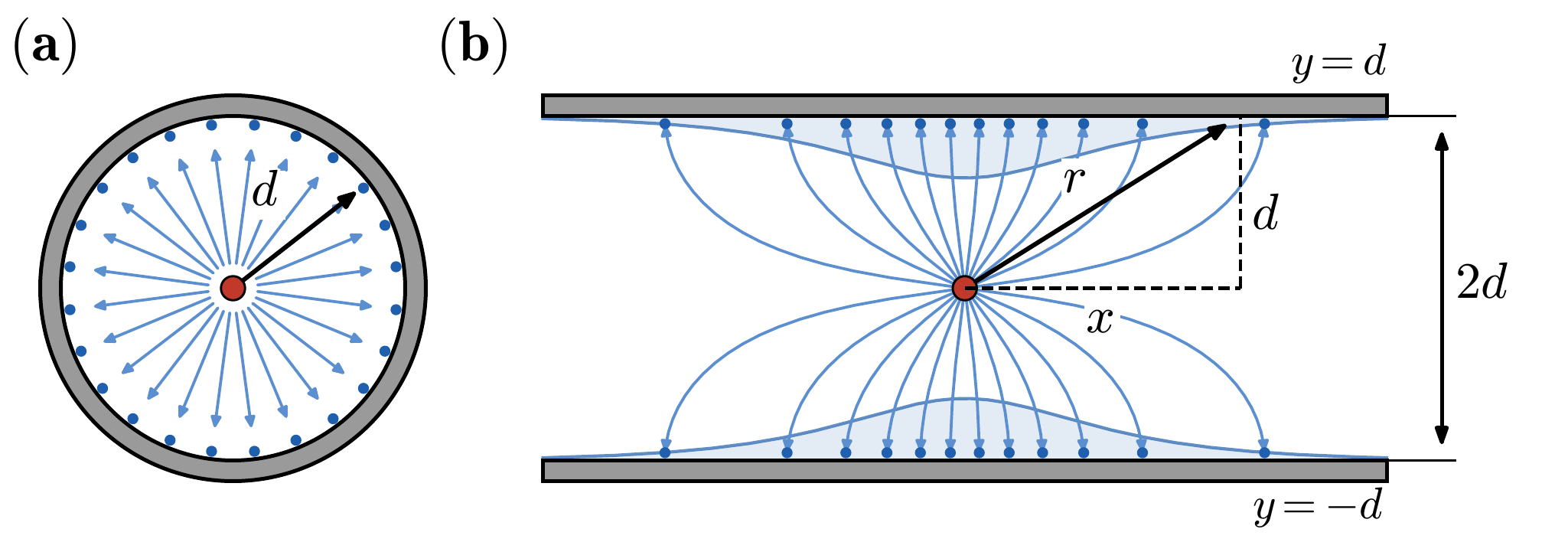}
   \caption{Illustration of the image-charge distributions used in the
    Fresnel number interpretation. (a) Round pipe of radius $d$, the
    image charge is uniformly distributed around the circumference, giving
    $\langle x^2\rangle_{\mathrm{im}}=d^2/2$ for one transverse projection.
    (b) Two parallel plates at $y=\pm d$ in the wide plate limit, $W\gg d$. The image charge is concentrated on the two plates, giving $\langle x^2\rangle_{\mathrm{im}}=d^2$.}
    \label{fig:fresnel_image_charge}
\end{figure}

For a tapered collimator, the bunch spectrum extends up to
$\omega_\sigma=c/\sigma_s$~\cite{Lin2000}, so the relevant reduced
wavelength is $\lambda_\sigma/2\pi=c/\omega_\sigma=\sigma_s$; the
propagation distance is the taper length $L$; and the aperture is the
transverse footprint of the beam field where it meets the outer wall,
i.e., the induced image-charge distribution. We therefore introduce an
effective Fresnel number, defined per transverse projection,
\begin{equation}
N_F=\frac{\langle x^2\rangle_{\mathrm{im}}}{\sigma_s L},
\label{eq:fresnel_effective}
\end{equation}
where $\langle\cdots\rangle_{\mathrm{im}}$ is the average over the
image charge. Using the reduced wavelength gives
$N_F=2\pi N_F^{\mathrm{opt}}$; this constant factor is absorbed into the numerical coefficient of the taper law.
Figure~\ref{fig:fresnel_image_charge} illustrates the corresponding image-charge distributions.

For the round geometry, we approximate the image charge as uniformly distributed around the circular boundary at $r=d$. For one
transverse projection, $x=d\cos\theta$, and the corresponding mean-square transverse extent is
\begin{equation}
\left\langle x^2\right\rangle_{\mathrm{im}}^{\mathrm{rnd}}
=
\frac{d^2}{2\pi}
\int_0^{2\pi}\cos^2\theta\,d\theta
=
\frac{d^2}{2}.
\label{eq:x2_round}
\end{equation}
Using this effective transverse scale in the Fresnel parameter gives
\begin{equation}
N_F^{\mathrm{rnd}}
=
\frac{\left\langle x^2\right\rangle_{\mathrm{im}}^{\mathrm{rnd}}}
{\sigma_s L}
=
\frac{d^2}{2\sigma_s L}.
\label{eq:F_round}
\end{equation}

For the flat geometry with $W\gg d$, the image charge is distributed
over the plates at $y=\pm d$. The Green's function between two grounded parallel plates is given by the monopole potential in Eq.~(29) of Ref.~\cite{Bane2007},

\begin{equation}
\phi_m(x,y)=\ln\frac{\cosh u+\cos v}{\cosh u-\cos v},\qquad
u=\frac{\pi x}{2d},\quad v=\frac{\pi y}{2d}.
\end{equation}\\
On the upper plate, $y=d$ and therefore $v=\pi/2$. The normal derivative is

\begin{equation}
\left.
\frac{\partial\phi_m}{\partial y}
\right|_{y=d}
=
-\frac{\pi}{d}
\operatorname{sech}
\left(
\frac{\pi x}{2d}
\right),
\end{equation}\\
The corresponding induced surface charge density on each plate is
\begin{equation}
\Sigma(x)
=
-\frac{1}{4d}
\operatorname{sech}
\left(
\frac{\pi x}{2d}
\right),
\label{eq:flat_surface_charge}
\end{equation}
with$
\int_{-\infty}^{\infty}
\Sigma(x)\,dx
=
-{1}/{2}$. The transverse extent of this image-charge distribution is characterized by
its second moment,
\begin{equation}
\left\langle x^2 \right\rangle_{\mathrm{im}}^{\mathrm{flt}}
=
\frac{
\displaystyle
\int_{-\infty}^{\infty}
x^2\Sigma(x)\,dx
}{
\displaystyle
\int_{-\infty}^{\infty}
\Sigma(x)\,dx
}
=
\frac{
\displaystyle
\int_{-\infty}^{\infty}
x^2
\operatorname{sech}
\left(
\frac{\pi x}{2d}
\right)dx
}{
\displaystyle
\int_{-\infty}^{\infty}
\operatorname{sech}
\left(
\frac{\pi x}{2d}
\right)dx
}.
\label{eq:second_moment_flat}
\end{equation}
With the substitution
$t=\pi x/(2d)$, the denominator and nominator of Eq.~\eqref{eq:second_moment_flat} becomes
$\int_{-\infty}^{\infty}
\operatorname{sech}
\left(
\frac{\pi x}{2d}
\right)dx
=2d$, and
\begin{align}
\int_{-\infty}^{\infty}
x^2
\operatorname{sech}
\left(
\frac{\pi x}{2d}
\right)dx
&=
\left(
\frac{2d}{\pi}
\right)^3
\int_{-\infty}^{\infty}
t^2\operatorname{sech}(t)\,dt.
\end{align}

Using the standard integral $\int_{-\infty}^{\infty}
t^2\operatorname{sech}(t)\,dt
=
\frac{\pi^3}{4}$,
we get,$\int_{-\infty}^{\infty}
x^2
\operatorname{sech}
\left(
\frac{\pi x}{2d}
\right)dx
=2d^3$. 
Therefore, Eq.~\eqref{eq:second_moment_flat} becomes $\langle x^2\rangle_{\mathrm{im}}^{\mathrm{flt}}
=d^2$. The corresponding Fresnel number is
\begin{equation}
N_{F}^{\mathrm{flt}}=\frac{d^2}{\sigma_s L}.
\label{eq:F_flat}
\end{equation}
From Eq.~\eqref{eq:F_round} and Eq.~\eqref{eq:F_flat}, we obtain,
\begin{equation}
N_{F}^{\mathrm{flt}}=2N_{F}^{\mathrm{rnd}}.
\label{eq:F_flat_round}
\end{equation}

With the corresponding Fresnel number definitions, the arguments of the round and flat taper functions become $0.4N_F^{\mathrm{rnd}}$ and
$0.4N_F^{\mathrm{flt}}$, respectively. The taper law can therefore be written
as a single geometry-independent expression,
\begin{equation}
\frac{k_{\parallel,\mathrm{tpr}}^{\mathrm{opt}}(\sigma_s)}
{k_{\parallel,\mathrm{step}}(\sigma_s)}
\simeq
\left[\frac{2}{\pi}\tan^{-1}\!\left(0.4N_F\right)\right]^2,
\label{eq:fresnel_common_law}
\end{equation}
\\
where $N_F$ is the Fresnel number for the corresponding geometry and thus
provides a common physical basis for the axisymmetric and rectangular taper
scalings.

\section{Optical to Inductive Regime Transition}

\label{subsec:bunch_scales}

The Fresnel number provides a qualitative measure of the diffraction behavior of the tapered collimator. For $N_F\gg 1$, the loss factor approaches the step value, while for $N_F\leq1$, diffraction in the taper reduces it according to the arctan law from Eq.~\eqref{eq:fresnel_common_law}. Both limits correspond to the optical regime shown in Fig.~\ref{fig:norm_mu}. 
The onset of the inductive regime is set by the cutoff wavenumber, $\sigma_s>\sigma_i=1/(\sqrt{2\ln 2}\,k_c)$,
independently of $N_F$. At the transition,
\begin{equation}
N_F^{\mathrm{rnd}}(\sigma_i)
=
\frac{\sqrt{2\ln 2}\,j_{01}}{2}\frac{d}{L}
\simeq 1.42\frac{d}{L},
\label{eq:NF_round_transition}
\end{equation} 
and
\begin{equation}
N_F^{\mathrm{flt}}(\sigma_i)
\simeq
\frac{\sqrt{2\ln 2}\,\pi}{2}\frac{d}{L}
\simeq 1.85\frac{d}{L},
\label{eq:NF_flat_transition}
\end{equation}
so the inductive regime is reached at $N_F<1$ only for $d/L\lesssim 0.5$.

\begin{figure}[htbp]
    \centering
    \includegraphics[width=\columnwidth]{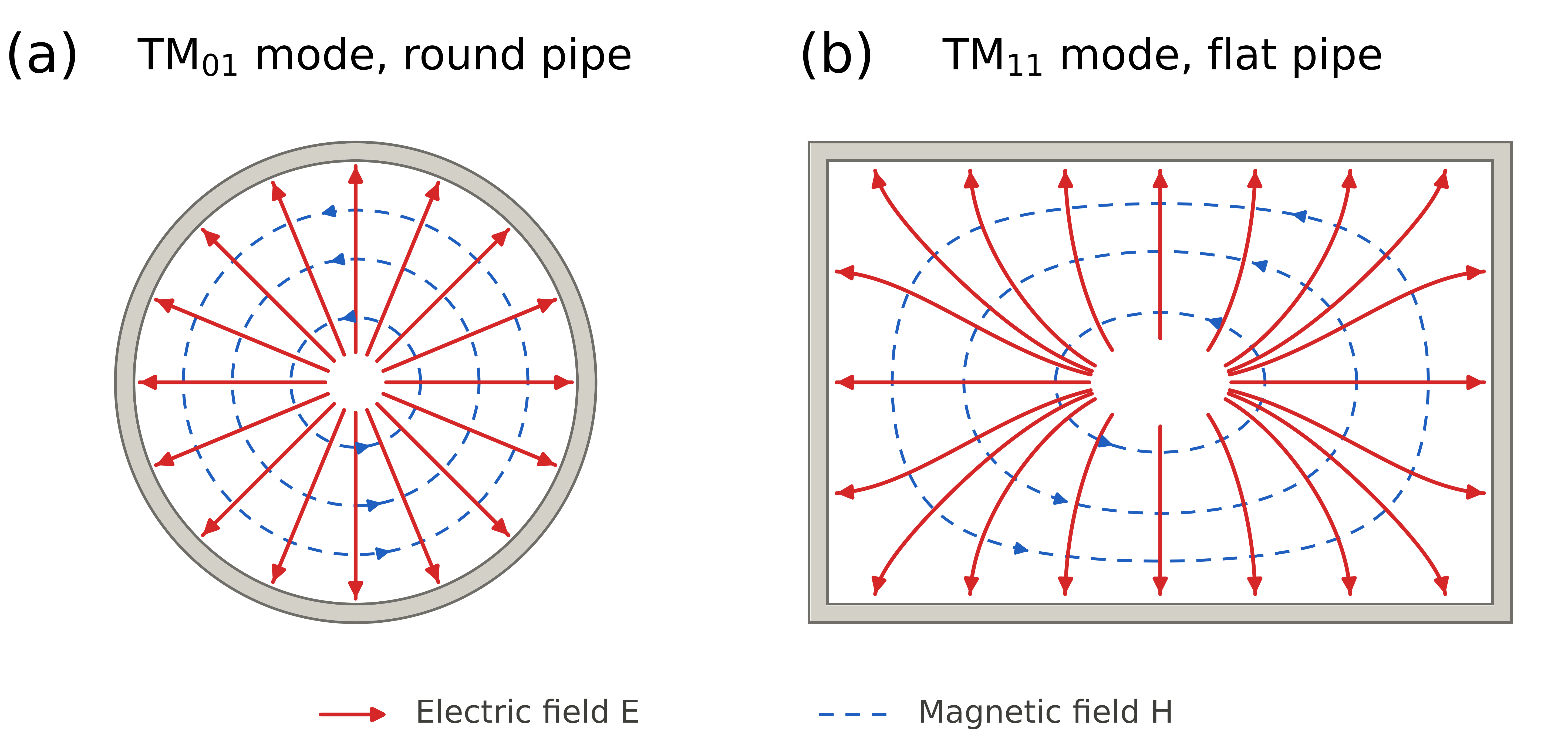}
   \caption{Electromagnetic field distributions in the pipe cross-section for the lowest-order TM modes: (a)~the TM$_{01}$ mode in a round geometry and (b)~the TM$_{11}$ mode in a flat geometry. Red solid and blue dashed arrows denote the electric field and magnetic field, receptively.}
    \label{fig:Cutoff_wavenumber}
\end{figure}

The transition from the optical to the inductive regime is characterized by the cutoff wavenumber of the lowest relevant longitudinal mode. Figure~\ref{fig:Cutoff_wavenumber}
illustrates the electromagnetic field distributions of the $\mathrm{TM}_{01}$ mode in a circular waveguide and the $\mathrm{TM}_{11}$ mode in a rectangular
waveguide. For the round geometry, the cutoff wavenumber of the $\mathrm{TM}_{01}$ mode is given by
\begin{equation}
k_c^{\mathrm{rnd}}
=
\frac{j_{01}}{d}
=
\frac{2.405}{d},
\label{eq:kc_round}
\end{equation}
where $j_{01}=2.405$ is the first zero of the Bessel function $J_0$.

To relate the modal scale to the rms bunch length, we use the half-maximum
point of a Gaussian distribution. For a Gaussian bunch,
$\lambda(z)/\lambda(0)=\exp[-z^2/(2\sigma_i^2)]$. At half maximum, $\lambda(z)/\lambda(0)=1/2$, which gives
$N = z/\sigma_i=\sqrt{2\ln 2}$.

For the axisymmetric collimator, the characteristic modal length is
$z_c^{\mathrm{rnd}}=1/k_c^{\mathrm{rnd}}$. Associating this
scale with the Gaussian half-maximum scale gives
$N\sigma_{\mathrm{i}}^{\mathrm{rnd}}=1/k_c^{\mathrm{rnd}}$. Therefore,
\begin{equation}
\sigma_{\mathrm{i}}^{\mathrm{rnd}}
=
\frac{1}{Nk_c^{\mathrm{rnd}}}
=
\frac{d}{N j_{01}}.
\label{eq:sigma_i_rnd}
\end{equation}

For the rectangular collimator, the cutoff wavenumber is written as
\begin{equation}
k_c^{\mathrm{flt}}
=
\pi
\sqrt{
\left(\frac{m}{W}\right)^2
+
\left(\frac{n}{2d}\right)^2
}.
\label{eq:kc_flat}
\end{equation}

We consider the $\mathrm{TM}_{11}$ mode of a rectangular waveguide with full width $W$ and full height $2d$, where $m=1$ and $n=1$. The corresponding modal length is
$z_c^{\mathrm{flt}}=1/k_c^{\mathrm{flt}}$. Using the same Gaussian
half-maximum condition, $N\sigma_{\mathrm{i}}^{\mathrm{flt}}
=1/k_c^{\mathrm{flt}}$, which gives
\begin{equation}
\sigma_{\mathrm{i}}^{\mathrm{flt}}
=
\frac{1}{N k_c^{\mathrm{flt}}}
=
\frac{1}
{N\pi
\sqrt{
\left(\dfrac{1}{W}\right)^2
+
\left(\dfrac{1}{2d}\right)^2
}}.
\label{eq:sigma_i_flat}
\end{equation}

Thus $\sigma_i\propto 1/k_c$ provides the characteristic scale separating the optical and inductive branches of the tapered collimator.

\section{Two-Regime Loss Factor Model of Tapered Collimators}

We first consider the corresponding untapered step collimators, which
provide the reference loss factors for the axisymmetric and rectangular
geometries. We then extend the analysis to tapered collimators and consider two bunch-length regimes: optical and inductive.

\subsection{Optical Regime}

\subsubsection{Step}

For an axisymmetric step collimator, the longitudinal impedance is given by~\cite{Balakin1983,Bisognano1988,Heifets1991}
\begin{equation}
Z_{\parallel,\mathrm{step}}^{\mathrm{rnd}}
=
\frac{4}{c}
\ln\left(\frac{d}{b}\right),
\label{eq:zstep_round}
\end{equation}
where $b$ and $d$ are the smaller and larger radii, respectively.
For a Gaussian bunch with rms length $\sigma_s$, the corresponding
loss factor is

\begin{equation}
k_{\parallel,\mathrm{step}}^{\mathrm{rnd}}
=
\frac{c}{2\sqrt{\pi}\sigma_s}
Z_{\parallel,\mathrm{step}}^{\mathrm{rnd}} .
\label{eq:kstep_round}
\end{equation}

For the rectangular collimator, the step-transition impedance can be
obtained from the optical approximation for a flat aperture
\cite{Bane2007}. In the wide-flat limit, $W\gg d$, the derivation begins with the monopole potential given in Eq. (29) of Ref.~\cite{Bane2007} for two parallel conducting plates,
\begin{equation}
\phi_m(x,y)
=
\ln
\left[
\frac{
\cosh\left(\dfrac{\pi x}{2d}\right)
+
\cos\left(\dfrac{\pi y}{2d}\right)
}{
\cosh\left(\dfrac{\pi x}{2d}\right)
-
\cos\left(\dfrac{\pi y}{2d}\right)
}
\right].
\label{eq:phi_flat}
\end{equation}
At the smaller aperture, $y=b$, and using $\alpha=b/d$, and differentiating Eq.~(\ref{eq:phi_flat}) with respect to $y$ gives
\begin{equation}
\frac{\partial \phi_m}{\partial y}(x,b)
=
-\frac{2\pi}{d}
\frac{
\cosh\left(\dfrac{\pi x}{2d}\right)
\sin\left(\dfrac{\pi\alpha}{2}\right)
}{
\cosh\left(\dfrac{\pi x}{d}\right)
-
\cos(\pi\alpha)
}.
\label{eq:dphi_at_b}
\end{equation}
Using Eq.~\eqref{eq:phi_flat}, the corresponding potential at $y=b$ is

\begin{equation}
\phi_m(x,b)
=
\ln
\left[
\frac{
\cosh\left(\dfrac{\pi x}{2d}\right)
+
\cos\left(\dfrac{\pi\alpha}{2}\right)
}{
\cosh\left(\dfrac{\pi x}{2d}\right)
-
\cos\left(\dfrac{\pi\alpha}{2}\right)
}
\right].
\label{eq:phi_at_b}
\end{equation}
The longitudinal impedance can then be written as
\begin{equation}
Z_{\parallel,\mathrm{step}}^{\mathrm{flt}}
=
-\frac{2}{\pi c}
\int_{0}^{\infty}
\phi_m(x,b)
\frac{\partial\phi_m}{\partial y}(x,b)
\,dx .
\label{eq:zflat_general}
\end{equation}
Substituting Eqs.~(\ref{eq:dphi_at_b}) and (\ref{eq:phi_at_b}) into Eq.~(\ref{eq:zflat_general}) gives
\begin{align}
Z_{\parallel,\mathrm{step}}^{\mathrm{flt}}
&=
\frac{4}{dc}
\int_{0}^{\infty}
dx\,
\frac{
\sin\left(\dfrac{\pi\alpha}{2}\right)
\cosh\left(\dfrac{\pi x}{2d}\right)
}{
\cos(\pi\alpha)
-
\cosh\left(\dfrac{\pi x}{d}\right)
}
\nonumber\\
&\qquad\times
\ln
\left[
\frac{
\cosh\left(\dfrac{\pi x}{2d}\right)
-
\cos\left(\dfrac{\pi\alpha}{2}\right)
}{
\cosh\left(\dfrac{\pi x}{2d}\right)
+
\cos\left(\dfrac{\pi\alpha}{2}\right)
}
\right].
\label{eq:zflat_dimensional}
\end{align}
Introducing the dimensionless coordinate \(x'=x/(2d)\), we obtain
\begin{equation}
\begin{aligned}
Z_{\parallel,\mathrm{step}}^{\mathrm{flt}}
=
\frac{8}{c}
\int_{0}^{\infty}
dx'\,
&
\frac{
\sin\left(\dfrac{\pi\alpha}{2}\right)
\cosh(\pi x')
}{
\cos(\pi\alpha)-\cosh(2\pi x')
}
\\
&\times
\ln
\left[
\frac{
\cosh(\pi x')
-
\cos\left(\dfrac{\pi\alpha}{2}\right)
}{
\cosh(\pi x')
+
\cos\left(\dfrac{\pi\alpha}{2}\right)
}
\right].
\end{aligned}
\label{eq:zstep_flat}
\end{equation}
Therefore, the loss factor of the flat collimator is
\begin{equation}
k_{\parallel,\mathrm{step}}^{\mathrm{flt}}
=
\frac{c}{2\sqrt{\pi}\sigma_s}
Z_{\parallel,\mathrm{step}}^{\mathrm{flt}}.
\label{eq:kstep_flat}
\end{equation}

The corresponding mks unit is obtained by multiplying by \(cZ_0/4\pi\), where \(Z_0 \approx 376.73~\Omega\) is the impedance of free space.

\subsubsection{Tapered}

For short bunches, tapering reduces the loss factor relative to the
corresponding step collimator. Wakefield simulations with ECHO and
GdfidL show that this reduction can be represented by a
geometry-dependent taper factor. For axisymmetric collimators with $d-b>b$, ECHO simulations give
$\mu_{\mathrm{rnd}}=0.2$~\cite{Blednykh2010}, whereas for wide
rectangular collimators with $d-b>b$ and $W\gg d$, GdfidL simulations give $\mu_{\mathrm{flt}}=0.4$~\cite{Blednykh2011}, consistent with Eq.~\eqref{eq:F_flat_round}.

The optical regime loss factor is
\begin{equation}
k_{\parallel,\mathrm{tpr}}^{\mathrm{opt}}(\sigma_s)
=
k_{\parallel,\mathrm{step}}(\sigma_s)
\left[
\frac{2}{\pi}
\tan^{-1}\!\left(
\frac{\mu d^2}{\sigma_s L}
\right)
\right]^2.
\label{eq:kopt}
\end{equation}
The optical regime is identified by
\begin{equation}
\sigma_s \leq \sigma_i ,
\end{equation}
where $\sigma_i$ denotes the characteristic bunch length separating
the optical and inductive regimes.

\subsection{Inductive Regime}

As the bunch length increases beyond $\sigma_i$, the bunch spectrum
shifts toward lower frequencies and the loss factor departs from the
short-bunch optical behavior. The following derivation is based on an
asymptotic lowest mode approximation, in which the long bunch response
is assumed to be dominated by the lowest relevant electromagnetic mode,
characterized by the cutoff wavenumber $k_c$.

For a Gaussian bunch of rms length $\sigma_s$, the loss
factor can be expressed in the wavenumber domain as
\begin{equation}
k_{\parallel}(\sigma_s)
=
\frac{c}{\pi}
\int_0^{\infty}
\operatorname{Re}\!\left[Z_{\parallel}(k)\right]
\exp\!\left(-k^2\sigma_s^2\right)\,dk .
\label{eq:loss_spectral}
\end{equation}
The factor $\exp(-k^2\sigma_s^2)$ represents the Gaussian spectral
weighting of the bunch and increasingly suppresses high wavenumber
components as $\sigma_s$ increases.

Within the asymptotic lowest mode approximation,
$\operatorname{Re}[Z_{\parallel}(k)]$ is taken to vary slowly near the
lowest cutoff. The leading bunch-length dependence of the long bunch response can then be approximated by the Gaussian spectral tail above
$k_c$,
\begin{equation}
k_{\parallel,\mathrm{tpr}}^{\mathrm{ind}}(\sigma_s)
\propto
\int_{k_c}^{\infty}
\exp\!\left(-k^2\sigma_s^2\right)\,dk .
\label{eq:cutoff_integral}
\end{equation}
Here, $u=k\sigma_s$, with $dk=du/\sigma_s$, and using the definition of the complementary
error function~\cite{Olver2010,DLMF},
$\operatorname{erfc}(x)=\frac{2}{\sqrt{\pi}}\int_x^{\infty}e^{-t^2}\,dt$. Eq.~\eqref{eq:cutoff_integral} becomes
\begin{equation}
k_{\parallel,\mathrm{tpr}}^{\mathrm{ind}}(\sigma_s)
\propto
\frac{\sqrt{\pi}}{2\sigma_s}
\operatorname{erfc}\!\left(k_c\sigma_s\right),
\label{eq:erfc_inductive}
\end{equation}
For sufficiently large $k_c\sigma_s$, the complementar error function
has the asymptotic form
\begin{equation}
\operatorname{erfc}(x)
\simeq
\frac{\exp(-x^2)}
{\sqrt{\pi}\,x},
\label{eq:erfc_asymptotic}
\end{equation}
Substituting Eq.~\eqref{eq:erfc_asymptotic} into
Eq.~\eqref{eq:erfc_inductive} gives the long-bunch scaling 
\begin{equation}
k_{\parallel,\mathrm{tpr}}^{\mathrm{ind}}(\sigma_s)
\propto
\frac{1}{k_c\sigma_s^2}
\exp\!\left(-k_c^2\sigma_s^2\right).
\label{eq:inductive_asymptotic}
\end{equation}
Normalizing Eq.~\eqref{eq:inductive_asymptotic} to its value at the characteristic bunch length $\sigma_i$ for $\sigma_s>\sigma_i$ gives

\begin{equation}
\frac{k_{\parallel,\mathrm{tpr}}^{\mathrm{ind}}(\sigma_s)}
{k_{\parallel,\mathrm{tpr}}^{\mathrm{ind}}(\sigma_i)}
\simeq
\left(
\frac{\sigma_i}{\sigma_s}
\right)^2
\exp\!\left[
-k_c^2
\left(
\sigma_s^2-\sigma_i^2
\right)
\right],
\label{eq:inductive_absolute_ratio}
\end{equation}
\\
Matching the inductive branch continuously to the optical
branch at $\sigma_s=\sigma_i$, such that
$k_{\parallel,\mathrm{tpr}}^{\mathrm{ind}}(\sigma_i)
= k_{\parallel,\mathrm{tpr}}^{\mathrm{opt}}(\sigma_i)$, the inductive loss factor is expressed as
\begin{equation}
k_{\parallel,\mathrm{tpr}}^{\mathrm{ind}}(\sigma_s)
=
k_{\parallel,\mathrm{tpr}}^{\mathrm{opt}}(\sigma_i)
\left(
\frac{\sigma_i}{\sigma_s}
\right)^2
\exp\!\left[
-k_c^2
\left(
\sigma_s^2-\sigma_i^2
\right)
\right].
\label{eq:inductive}
\end{equation}
\\
Using, Eq.~\eqref{eq:kopt}, the absolute loss factor from Eq.~\eqref{eq:inductive} can be writtern as 
\begin{equation}
\begin{aligned}
 k_{\parallel,\mathrm{tpr}}^{\mathrm{ind}}(\sigma_s)
&=
k_{\parallel,\mathrm{step}}(\sigma_i)
\left[
\frac{2}{\pi}
\tan^{-1}\!\left(
\frac{\mu d^2}{\sigma_i L}
\right)
\right]^2
\left(
\frac{\sigma_i}{\sigma_s}
\right)^2
\\
&\qquad\qquad\qquad\quad \times
\exp\!\left[
-k_c^2
\left(
\sigma_s^2-\sigma_i^2
\right)
\right].
\end{aligned}
\label{eq:inductive1}
\end{equation}
which is identical to Eq.~\eqref{eq:two_regime_intro}.

From Eq.~\eqref{eq:kstep_round} and ~\eqref{eq:kstep_flat}, the step-transition loss factor is
$k_{\parallel,\mathrm{step}}\propto1/\sigma_s$, therefore
\begin{equation}
\frac{k_{\parallel,\mathrm{step}}(\sigma_i)}
{k_{\parallel,\mathrm{step}}(\sigma_s)}
=
\frac{\sigma_s}{\sigma_i}.
\label{eq:step_ratio}
\end{equation}
\\
From Eqs.~\eqref{eq:inductive_absolute_ratio} and
\eqref{eq:step_ratio}, we get
\begin{equation}
\frac{
k_{\parallel,\mathrm{tpr}}^{\mathrm{ind}}(\sigma_s)
}{
k_{\parallel,\mathrm{step}}(\sigma_s)
}
=
\frac{
k_{\parallel,\mathrm{tpr}}^{\mathrm{opt}}(\sigma_i)
}{
k_{\parallel,\mathrm{step}}(\sigma_i)
}
\left(
\frac{\sigma_i}{\sigma_s}
\right)
\exp\!\left[
-k_c^2
\left(
\sigma_s^2-\sigma_i^2
\right)
\right].
\label{eq:normalized_inductive_ratio}
\end{equation}
\\
At the transition point, the optical branch of Eq.~\eqref{eq:kopt} gives
$k_{\parallel,\mathrm{tpr}}^{\mathrm{opt}}(\sigma_i)/k_{\parallel,\mathrm{step}}(\sigma_i)
=
\left[
\frac{2}{\pi}
\tan^{-1}\!\left(
\frac{\mu d^2}{\sigma_i L}
\right)
\right]^2$.
Hence, for $\sigma_s>\sigma_i$, the inductive branch of the normalized two-regime model is
\begin{equation}
\begin{aligned}
\frac{k_{\parallel,\mathrm{tpr}}^{\mathrm{ind}}(\sigma_s)}
{k_{\parallel,\mathrm{step}}(\sigma_s)}
&=
\left[
\frac{2}{\pi}
\tan^{-1}\!\left(
\frac{\mu d^2}{\sigma_i L}
\right)
\right]^2
\left(
\frac{\sigma_i}{\sigma_s}
\right)
\\
&\qquad\qquad\quad\times
\exp\!\left[
-k_c^2
\left(
\sigma_s^2-\sigma_i^2
\right)
\right].
\end{aligned}
\label{eq:normalized_inductive}
\end{equation}

\subsection{Combined Optical–Inductive Model}

Using the optical result of Eq.~\eqref{eq:kopt} together with the
long-bunch result of Eq.~\eqref{eq:inductive}, the absolute loss factor
of the tapered structure can be written as

\begin{equation}
k_{\parallel,\mathrm{tpr}}(\sigma_s)
=
\begin{cases}
\displaystyle
k_{\parallel,\mathrm{tpr}}^{\mathrm{opt}}(\sigma_s),
&
\begin{array}{l}
\sigma_s\leq\sigma_i\\
\text{(optical)}
\end{array}
\\[8pt]
\displaystyle
\begin{aligned}
&k_{\parallel,\mathrm{tpr}}^{\mathrm{opt}}(\sigma_i)
\left(\frac{\sigma_i}{\sigma_s}\right)^2
\\[-2pt]
&\quad\times
\exp\!\left[
-k_c^2\left(\sigma_s^2-\sigma_i^2\right)
\right],
\end{aligned}
&
\begin{array}{l}
\sigma_s>\sigma_i\\
\text{(inductive)}
\end{array}
\end{cases}
\label{eq:complete_model}
\end{equation}
\\
The two branches are continuous at $\sigma_s=\sigma_i$. The absolute
inductive branch contains $(\sigma_i/\sigma_s)^2$, whereas the
corresponding normalized form introduced in
Eq.~\eqref{eq:normalized_inductive} contains $\sigma_i/\sigma_s$ because
$k_{\parallel,\mathrm{step}}\propto1/\sigma_s$. This formulation connects the short-bunch optical behavior to the long-bunch inductive regime within a single loss factor model.

\subsection{Frequency and Catch-Up Distance Conditions}

The conventional optical approximation is associated with two main
conditions: the bunch spectrum must extend to sufficiently high
frequencies, and the transition length must be short compared with
the catch-up distance \cite{Bane2007}. We examine both conditions
for the tapered collimators considered here.

The taper angle is written as $\theta_{\mathrm{taper}}
= {(d-b)}/{L}$~\cite{Blednykh2012,Blednykh2013}. The corresponding characteristic frequency is
\begin{equation}
f_{\mathrm{taper}}
=
\frac{c}
{2\pi b\theta_{\mathrm{taper}}}.
\label{eq:taper_frequency}
\end{equation}
For $b=4.6~\mathrm{mm}$, $d=18~\mathrm{mm}$, and $L=300~\mathrm{mm}$, $f_{\mathrm{taper}}\simeq 232~\mathrm{GHz}$; thus, the high frequency condition for the optical approximation requires frequencies well above this characteristic scale.

The second condition is determined by the catch-up distance that is used to determine whether the electromagnetic field generated at transition can interact with the bunch within the tapered region. For a Gaussian bunch with characteristic angular frequency \cite{Siegman1986} is 
$\omega \simeq c/{\sigma_s}$, the catch-up distance can be written as \cite{Stupakov2009}
\begin{equation}
L_{\mathrm{catch}}
=
\frac{b^2\omega}{c}
=
\frac{b^2}{\sigma_s}.
\label{eq:catchup}
\end{equation}
The conventional short transition condition is
$L<L_{\mathrm{catch}}$~\cite{Bane2007}. For sufficiently short bunches,
$L_{\mathrm{catch}}$ is large and this condition is satisfied. As the
bunch length increases, $L_{\mathrm{catch}}\propto1/\sigma_s$ decreases
and can become smaller than the taper length.

Using the Fresnel number scaling introduced above,
$N_F\sim d^2/(\sigma_s L)$, together with
$L_{\mathrm{catch}}\sim b^2/\sigma_s$, the ratio of the
catch-up distance to the taper length can be expressed as
\begin{equation}
\frac{L_{\mathrm{catch}}}{L}
\sim
N_F\left(\frac{b}{d}\right)^2 .
\label{eq:catchup_ratio}
\end{equation}
Near $N_F\sim1$, this ratio therefore scales as $(b/d)^2$.
For a sufficiently large change in aperture, such as $d-b>b$~\cite{Blednykh2011}, corresponding to $d>2b$, the ratio is reduced accordingly, and the catch-up distance can become shorter than the taper length, $L_{\mathrm{catch}}<L$.

The wakefield simulations nevertheless show that the loss factor can continue to follow the short bunch optical scaling even when $L_{\mathrm{catch}}<L$. Thus, the conventional catch-up distance condition alone does not provide a sharp boundary for the optical regime. This behavior is consistent with the Fresnel number interpretation discussed above. In the present model, the cutoff wavenumber, $k_c$ determines the
characteristic bunch length separating the optical and inductive branches, with $\sigma_i\propto 1/k_c$.

\section{Results and Discussion}
Unless otherwise stated, the reference parameters used in the parametric
analysis are listed in Table~\ref{tab:reference_parameters}. When studying
the dependence on a particular parameter, that parameter is varied while
all other parameters are kept at their reference values.

\begin{table}[H]
\centering
\caption{Reference parameters and characteristic bunch-length scales.}
\label{tab:reference_parameters}
\begin{tabular}{lcc}
\hline
Parameter & Round & Flat \\
\hline
Small radius/aperture, $b$                     & $4.6$ mm  & $4.6$ mm \\
Large radius/aperture, $d$                     & $18$ mm   & $18$ mm \\
Taper length, $L$                       & $300$ mm  & $300$ mm \\
Gap, $g$                                & $100$ mm  & $100$ mm \\
Chamber width, $W$                      & --        & $67$ mm \\
Optical--inductive transition, $\sigma_i$
                                         & $6.34$ mm & $8.55$ mm \\
\hline
\end{tabular}
\end{table}

\subsection{Longitudinal Wakefield Simulation}

We performed numerical simulations using both ECHO and GdfidL for the flat tapered collimator. For the flat
geometry, $b$ and $d$ denote the smaller and larger vertical half-apertures, respectively, while $W$ is the full chamber width. The longitudinal wake potential $W_{\parallel}(s)$ is calculated for a
Gaussian bunch with rms length $\sigma_s$.

For a finite bunch, the longitudinal wake potential is obtained by convolving the point charge wake function $w_{\parallel}(s)$ with as
$W_{\parallel}(s)
=
\int_{-\infty}^{\infty}
w_{\parallel}(s-s')\lambda(s')\,ds'$, where $\lambda(s)$ is the normalized longitudinal distribution with $\int_{-\infty}^{\infty}\lambda(s)\,ds=1$. We obtain $W_{\parallel}(s)$ directly from wakefield simulations for the specified Gaussian bunch. The corresponding longitudinal loss factor is then evaluated in the wake domain as
$k_{\parallel}
=
-\int_{-\infty}^{\infty}
W_{\parallel}(s)\lambda(s)\,ds$,
where the sign convention is chosen such that $k_{\parallel}>0$
corresponds to energy loss by the bunch.

\begin{figure}[H]
    \centering
    \includegraphics[width=1\linewidth]{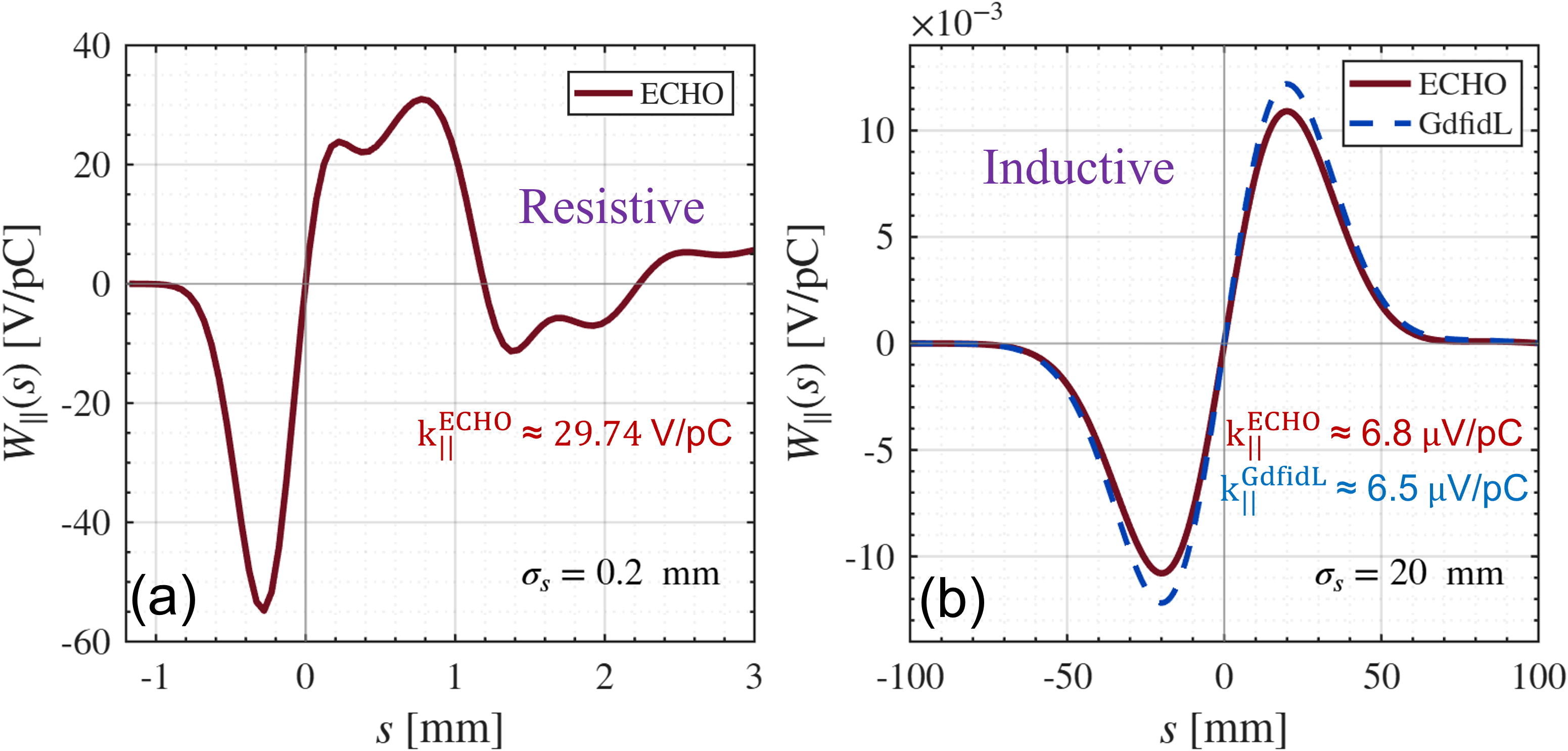}
     \caption{Simulated longitudinal wakefields for two bunch lengths:
     (a) short bunch ($\sigma_s=0.2$ mm), exhibiting a
     predominantly resistive response obtained using ECHO; and
     (b) long bunch ($\sigma_s=20$ mm), exhibiting a
     predominantly inductive response obtained using
ECHO and GdfidL..}
    \label{fig:flat_wakefield}
\end{figure}

Figure~\ref{fig:flat_wakefield} shows the longitudinal wakefields of the flat collimator simulated with ECHO and GdfidL. Two bunch lengths are selected to illustrate the resistive and inductive responses~\cite{Morton1993}. For the short bunch ($\sigma_s=0.2$~mm), the wakefield calculated with ECHO exhibits oscillations and a loss factor of $k_{\parallel,\mathrm{tpr}}=29.74$~V/pC, indicating a predominantly resistive response. 
In contrast, for the long bunch ($\sigma_s=20$~mm), the wakefield shows a nearly antisymmetric bipolar profile, with a substantially reduced loss factors of $k_{\parallel,\mathrm{tpr}}\approx6.8$~$\mu$V/pC and $6.5$~$\mu$V/pC obtained from ECHO and GdfidL, respectively, indicating a predominantly inductive response with negligible net energy loss. 
The small difference between the numerical results may arise from the
chosen step sizes of both codes, and the GdfidL results, obtained with
a $50~\mu$m step size, are expected to converge toward the ECHO results
as the step size of GdfidL is further reduced.

\subsection{Normalized Loss Factor Scaling}

We present the numerical results for the round and flat tapered collimators in normalized form to examine the geometrical scaling and directly compare them with the analytical model in the optical and inductive regimes.

\begin{figure}[H]
    \centering
    \includegraphics[width=\columnwidth]{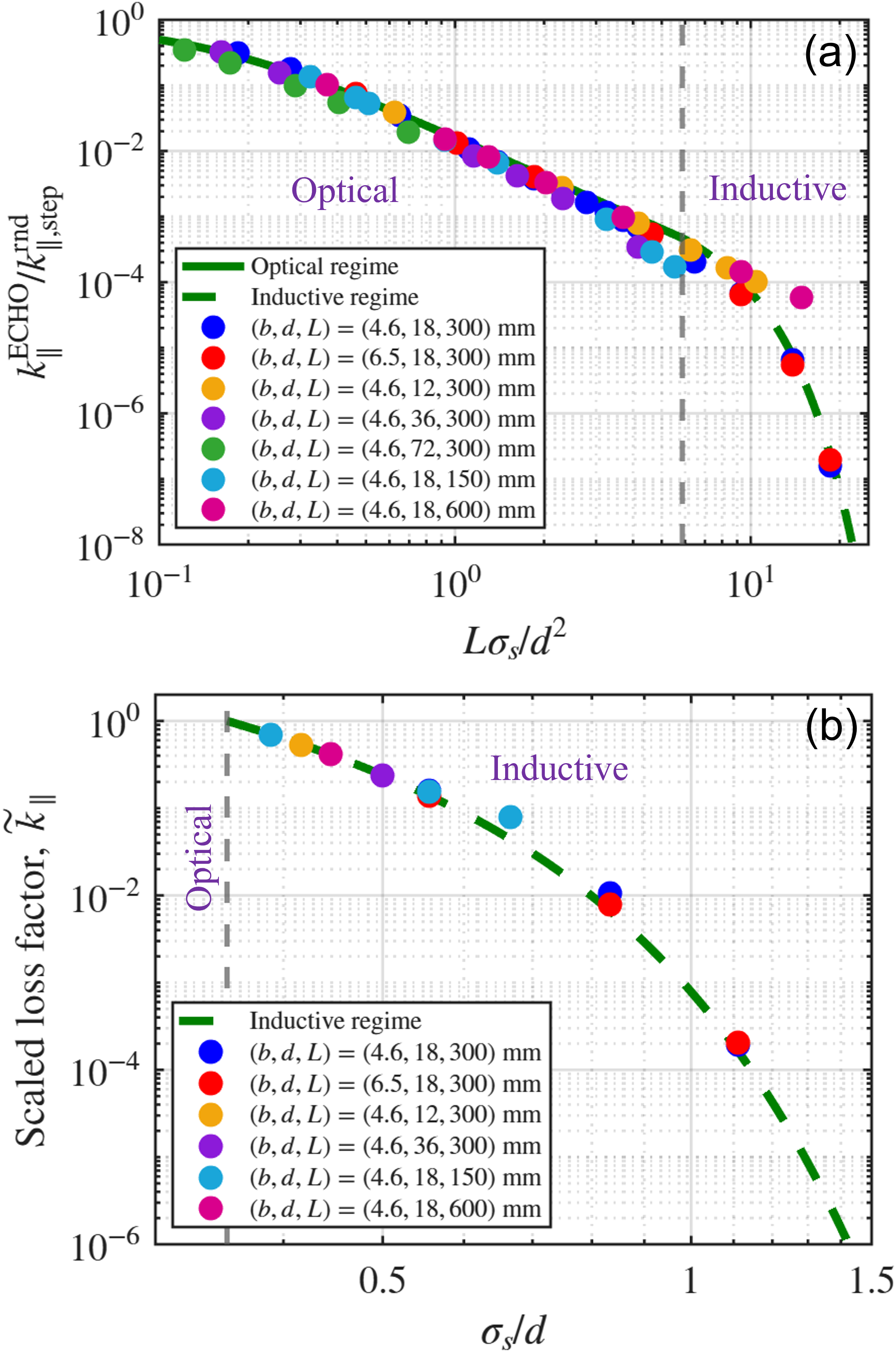}
    \caption{Normalized loss factor of the round tapered collimator:
    (a) optical regime and (b) inductive regime. The symbols represent
    numerical results obtained using ECHO. The vertical dashed lines indicate the optical-inductive transition $\sigma_i^{\mathrm{rnd}}$.}
\label{fig:rnd_normalized}
\end{figure}

For the round tapered collimator, the reference geometry is $b=4.6$ mm, $d=18$ mm, and $L=300$ mm. Additional geometries are considered by individually varying $b$ to $6.5$ mm, $d$ to $12$, $36$, and $72$ mm, and $L$ to $150$ and $600$ mm, respectively, while keeping the remaining parameters fixed. In
Fig.~\ref{fig:rnd_normalized}(a), the numerical loss factor is normalized by the step loss factor of Eq.~\eqref{eq:kstep_round} as $k_{\parallel}^{\mathrm{ECHO}}/
k_{\parallel,\mathrm{step}}^{\mathrm{rnd}}$ and plotted against the scaled bunch length
$L\sigma_s/d^2$. The analytical data from Eq.~\eqref{eq:kopt} are plotted as the ratio of the tapered collimator loss factor to the corresponding step loss factor
for different bunch lengths (solid green curve). The numerical results for the different geometries follow a common trend and agree well with the analytical predictions in the optical regime.

\begin{figure}[H]
    \centering
    \includegraphics[width=\columnwidth]{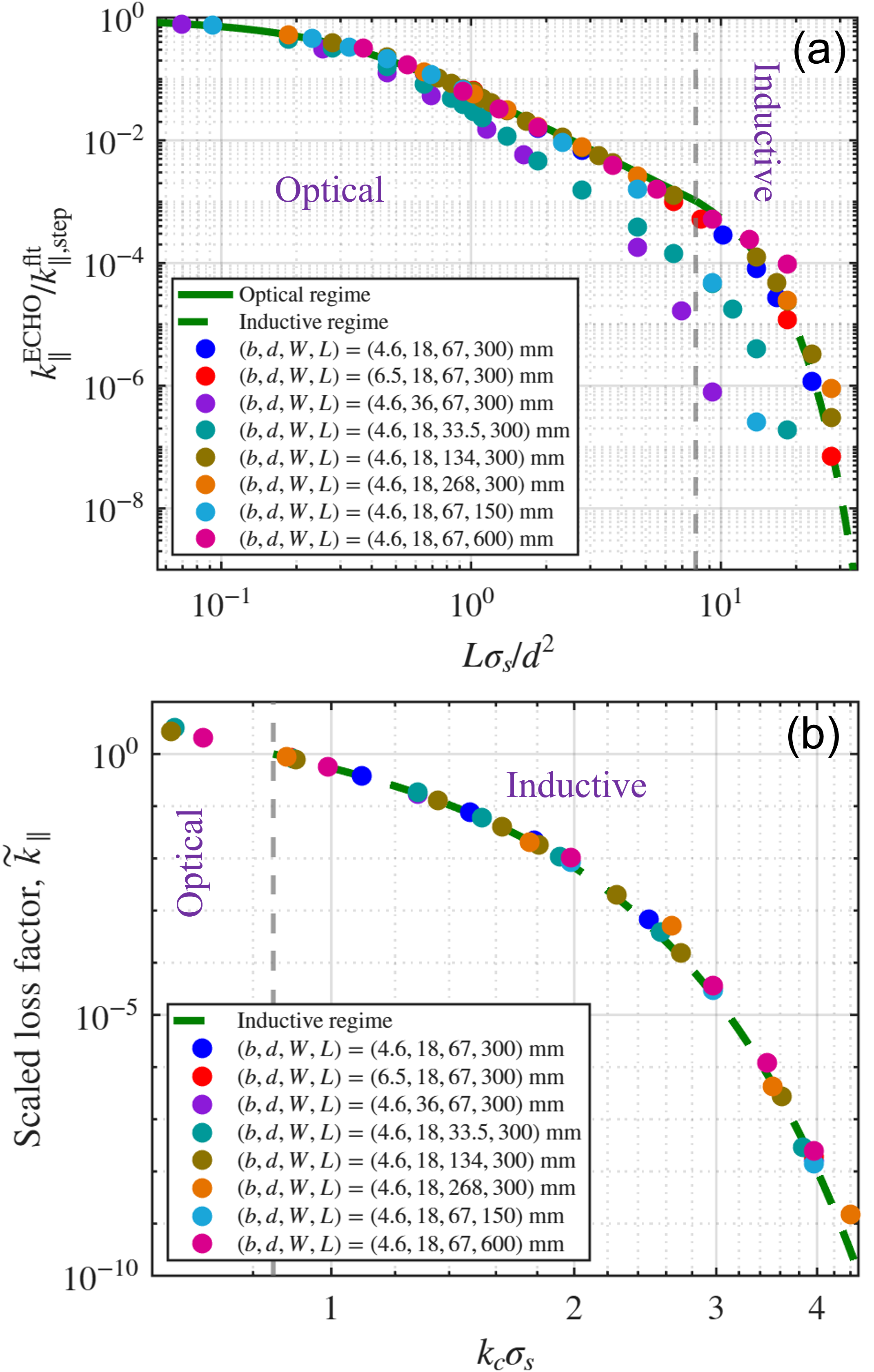}
    \caption{Normalized loss factor of the flat tapered collimator:
    (a) optical regime and (b) inductive regime. The symbols represent
    numerical results obtained using ECHO. The vertical dashed lines indicate the optical-inductive transition
    $\sigma_i^{\mathrm{flt}}$.}
    \label{fig:flat_normalized}
\end{figure}

Figure~\ref{fig:rnd_normalized}(b) shows the inductive regime results as a function of the scaled bunch length $\sigma_s/d$. The vertical axis represents the scaled loss factor
$\widetilde{k}_{\parallel}(\sigma_s)
=
[k_{\parallel}^{\mathrm{ECHO}}(\sigma_s)/
k_{\parallel}^{\mathrm{ECHO}}(\sigma_j)]
U_{\mathrm{rnd}}(r_j)$,
where $r=\sigma_s/d$, $\sigma_j$ is the first simulated bunch length
satisfying $\sigma_j\geq\sigma_i$, $r_j=\sigma_j/d$, $r_i=\sigma_i/d$. The analytical data is represented by
$U_{\mathrm{rnd}}(r)
=(r_i/r)^2
\exp[-j_{01}^{2}(r^2-r_i^2)]$
(dashed green curve).
The scaled numerical results follow the predicted inductive decay for longer bunches.

For the flat tapered collimator, we use the reference geometry
$b=4.6$~mm, $d=18$~mm, $W=67$~mm, and $L=300$~mm.
We apply the same variations in $b$, $d$, and $L$ as in the round
geometry and additionally vary the chamber width to
$W=33.5$, $134$, and $268$~mm, respectively. In Fig.~\ref{fig:flat_normalized}(a), the numerical loss factor is normalized by the step loss factor of Eq.~\eqref{eq:kstep_flat}
as $k_{\parallel}^{\mathrm{ECHO}}/
k_{\parallel,\mathrm{step}}^{\mathrm{flt}}$ and plotted against
$L\sigma_s/d^2$. 
The analytical predictions from Eq.~\eqref{eq:inductive1},
normalized by $k_{\parallel,\mathrm{step}}^{\mathrm{flt}}$,
are also plotted (solid green curve). The numerical results for the different geometries follow a common trend in the optical regime, while the influence of the finite chamber width becomes more pronounced toward the inductive regime. The larger deviations for $W=33.5$~mm, $d=18$~mm and
$W=67$~mm, $d=36$~mm arise from their transverse aspect ratios.
In both cases, $W/(2d)\approx1$, so the cross section approaches a
square rather than a wide flat rectangle. The corresponding transverse
image-charge distribution therefore differs from the wide flat limit underlying the analytical approximation, resulting in larger discrepancies between the analytical model and the numerical results.

To include the finite transverse scale, the inductive-regime results are
plotted in Fig.~\ref{fig:flat_normalized}(b) using $u=k_c\sigma_s$, where
$k_c=\pi\sqrt{1/W^2+1/(2d)^2}$. The vertical axis represents the scaled loss
factor
$\widetilde{k}_{\parallel}(\sigma_s)
=
[k_{\parallel}^{\mathrm{ECHO}}(\sigma_s)/
k_{\parallel}^{\mathrm{ECHO}}(\sigma_j)]
U_{\mathrm{flt}}(u_j)$,
where $\sigma_j$ is the first simulated bunch length satisfying
$\sigma_j\geq\sigma_i$, $u_j=k_c\sigma_j$, $u_i=k_c\sigma_i$.
The analytical results are presented by $U_{\mathrm{flt}}(u)
=
(u_i/u)^2\exp[-(u^2-u_i^2)]$ (dashed green curve).
As $k_c$ incorporates both the vertical half-aperture $d$ and the finite width $W$, the geometry dependence is captured by the scaled variable $u$. The vertical normalization further removes differences in the overall amplitude, allowing the different flat geometries to follow a common inductive trend.

\subsection{Parametric Analysis of Round and Flat Collimators}

\begin{figure}[H]
    \centering
    \includegraphics[width=1\columnwidth]{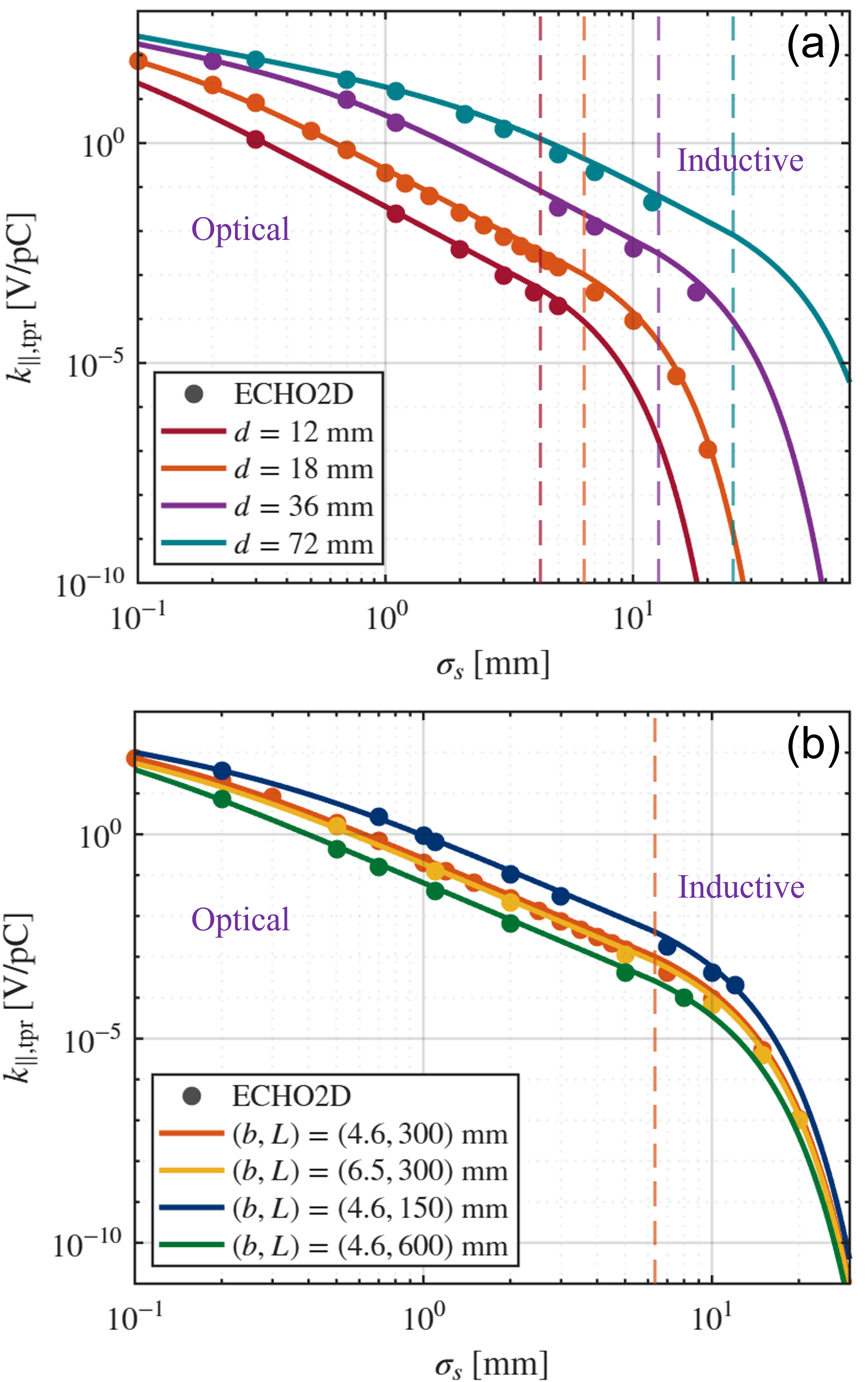}
    \caption{Dependence of the loss factor of the round tapered collimator on the geometrical parameters: (a) variation of $d$ for fixed $b=4.6$~mm and $L=300$~mm; (b) variation of $b$ and $L$ for fixed $d=18$~mm. The symbols represent numerical results obtained using ECHO. The vertical dashed lines indicate the optical-inductive transition $\sigma_i^{\mathrm{rnd}}$.}
    \label{fig:round_parameters}
\end{figure}

We systematically investigate the dependence of the loss factor and regime boundaries on $b$, $d$, and $L$ for round collimators, while also considering the chamber width $W$ for flat collimators. Figure~\ref{fig:round_parameters}(a) shows the dependence of the loss
factor of the round collimator on the large radius $d$ for fixed $b=4.6$~mm and $L=300$~mm. The transition bunch length scales linearly with the large radius, $\sigma_i^{\mathrm{rnd}}\propto d$. For
$d=12$, $18$, $36$, and $72$~mm, the corresponding transition bunch
lengths are $\sigma_i^{\mathrm{rnd}}\approx4.23$, $6.34$, $12.69$,
and $25.37$~mm, respectively. Thus, increasing $d$ shifts the
optical inductive transition toward larger bunch lengths. The
geometries considered here also satisfy the condition $d-b>b$~\cite{Blednykh2011}. Figure~\ref{fig:round_parameters}(b) shows the effects of the small
radius $b$ and taper length $L$ for fixed $d=18$~mm. Since
$\sigma_i^{\mathrm{rnd}}$ depends only on $d$, varying $b$ or $L$ does
not shift the transition, which remains at
$\sigma_i^{\mathrm{rnd}}\approx6.34$~mm. Increasing the taper length
reduces the loss factor, corresponding to weaker beam coupling for a
more gradual transition~\cite{Blednykh2011}.

\begin{figure}[H]
    \centering
    \includegraphics[width=\columnwidth]{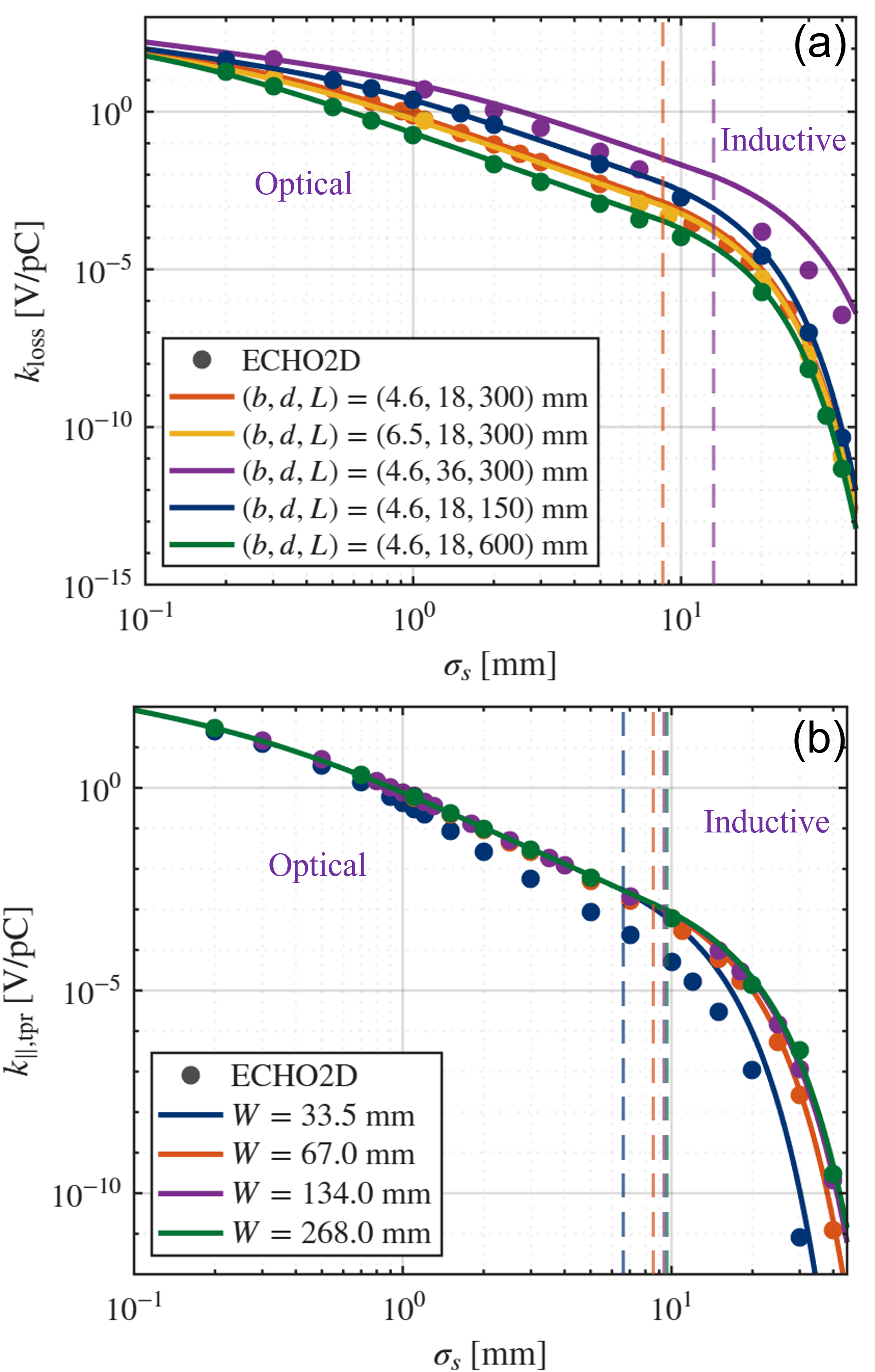}
    \caption{Dependence of the loss factor of the flat tapered collimator on the geometrical parameters: (a) variation of $b$, $d$, and
    $L$ for fixed chamber width $W=67$~mm; (b) variation of $W$ for fixed
    $b=4.6$~mm, $d=18$~mm, and $L=300$~mm. The symbols represent
    numerical results obtained using ECHO. The vertical dashed lines
    indicate the optical-inductive transition $\sigma_i^{\mathrm{flat}}$.}
    \label{fig:flat_parameters}
\end{figure}

Figure~\ref{fig:flat_parameters}(a) shows the dependence of the loss factor of flat collimator on $b$, $d$, and $L$ for a fixed chamber width $W=67$~mm. For $d=18$ and $36$~mm, the corresponding transition bunch lengths are
$\sigma_i^{\mathrm{flat}}\approx8.55$ and $13.23$~mm, respectively.
Thus, increasing $d$ shifts the optical inductive transition toward
longer bunch lengths. In contrast, varying $b$ or $L$ does not shift
$\sigma_i^{\mathrm{flat}}$, since the transition depends only on $d$ and
$W$. Increasing the taper length reduces the loss factor~\cite{Blednykh2011}, corresponding
to a more gradual geometrical transition. Figure~\ref{fig:flat_parameters}(b) shows the effect of the chamber width
$W$ for fixed $b=4.6$~mm, $d=18$~mm, and $L=300$~mm. For
$W=33.5$, $67$, $134$, and $268$~mm, the corresponding transition bunch
lengths are $\sigma_i^{\mathrm{flat}}\approx6.62$, $8.55$, $9.38$, and
$9.62$~mm, respectively. Increasing $W$ decreases the cutoff wavenumber
and shifts the transition toward longer bunch lengths.

The analytical flat chamber model is derived under the wide chamber
assumption, for which $W>2d$. The numerical results indicate that
quantitative agreement improves further when the chamber is sufficiently
wide that the side walls carry only a small fraction of the image charge.
This condition is approximately satisfied for
$W\gtrsim3.6d$, or equivalently $W/(2d)\gtrsim1.8$.
For narrower chambers, neglecting the finite width can overestimate the loss factor by up to an order of magnitude. This limitation is evident in Fig.~\ref{fig:flat_parameters}(a), where
$d=36$~mm gives $2d=72$~mm, exceeding the fixed width $W=67$~mm.
Similarly, in Fig.~\ref{fig:flat_parameters}(b), the case
$W=33.5$~mm is narrower than $2d=36$~mm for $d=18$~mm.
In both cases, the wide chamber assumption is violated, resulting in
larger discrepancies between the analytical model and the numerical results.
These deviations are therefore associated primarily with finite width
effects rather than with the optical to inductive transition scaling.

\subsection{Comparison of Axisymmetric and Flat Collimators}

\begin{figure}[H]
    \centering
    \includegraphics[width=\columnwidth]{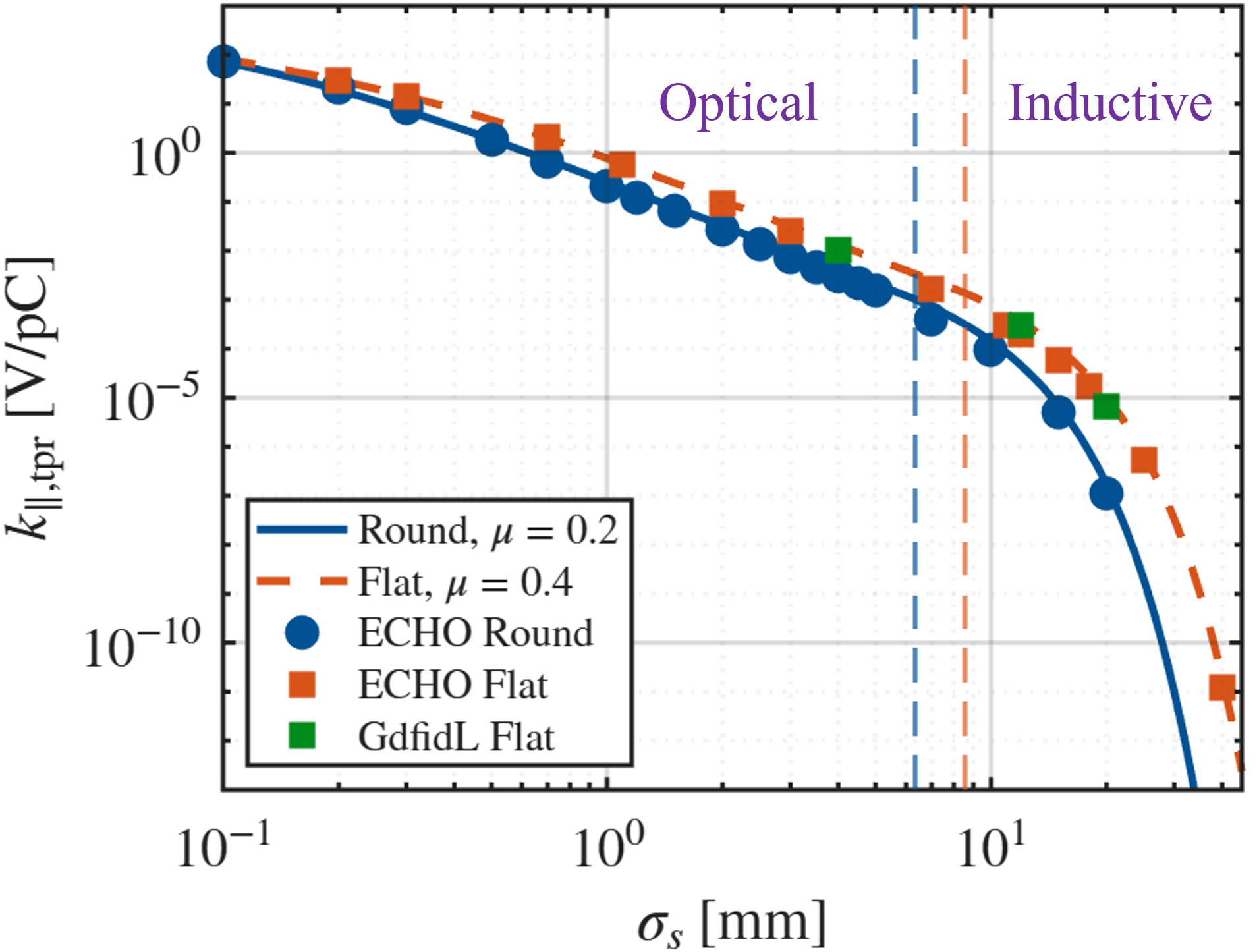}
    \caption{Comparison of the loss factors for round and flat tapered
    collimators. Symbols show the simulation results obtained with ECHO
    for round collimator and with ECHO and GdfidL for flat
    collimator. The vertical dashed lines
    indicate the inductive transition bunch lengths,
    $\sigma_i^{\mathrm{rnd}}=6.36$~mm and
    $\sigma_i^{\mathrm{flt}}=8.57$~mm.}
    \label{fig:round_flat_comparison}
\end{figure}

We compare the loss factors of the round and
flat tapered collimators for $b=4.6$~mm, $d=18$~mm, and
$L=300$~mm, with $W=67$~mm for the flat geometry shown in Fig.~\ref{fig:round_flat_comparison}.
The geometry dependent parameters $\mu_{\mathrm{rnd}}=0.2$
and $\mu_{\mathrm{flat}}=0.4$, originally introduced for the optical regime~\cite{Blednykh2010,Blednykh2011},
are retained in the combined optical-inductive regime model. The dependence of the taper correction on $d^2/(\sigma_s L)$ is
consistent with Fresnel number scaling (Eq.~\eqref{eq:F_flat_round}), supporting the use of these parameters across the optical to inductive transition.

At $\sigma_s=0.2$~mm, in the optical regime,
ECHO gives loss factors of 29.74 and 20.7~V/pC
for the flat and round collimators, respectively,
compared with 29.8 and 19.1~V/pC from Eq.~\eqref{eq:complete_model}. At $\sigma_s=20$~mm, in the inductive regime, the ECHO values decrease to
6.85 and 0.110~$\mu$V/pC for flat and round collimators, while the model predicts 10.1 and 0.167~$\mu$V/pC, respectively. For the flat structure at $\sigma_s=20$~mm, GdfidL gives 6.5~$\mu$V/pC, in good agreement with ECHO. These results demonstrate the rapid reduction in loss factor for longer bunches, with the model capturing the overall trend across both geometries.

\subsection{Lessons Learned: Detecting Geometry Errors in Simulation Input Models}

\begin{figure}[H]
    \centering
    \includegraphics[width=\columnwidth]{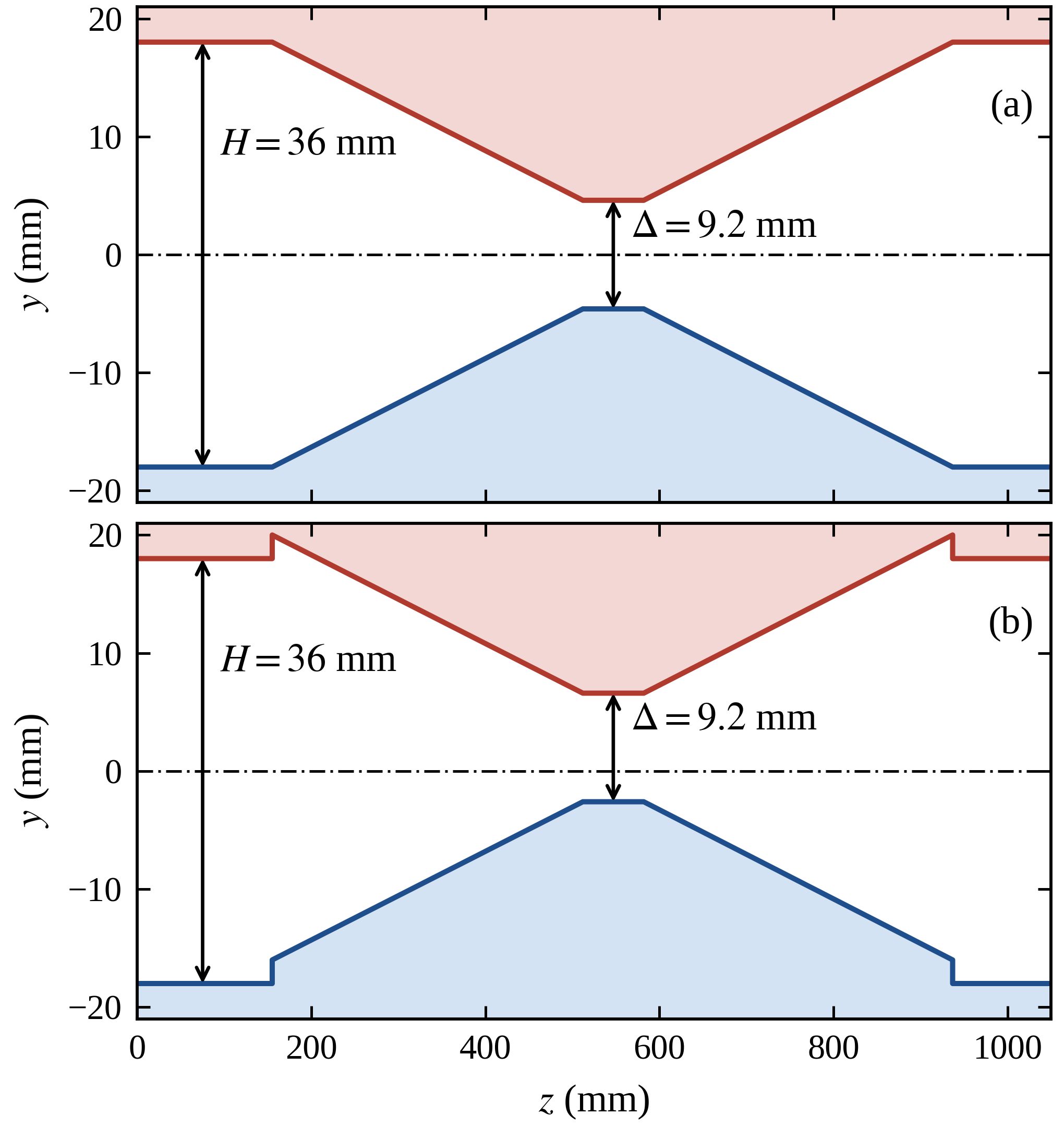}
    \caption{Geometry verification of the vertical collimator:
    (a) nominal symmetric jaw configuration and
    (b) vertically offset jaw configuration with reduced
    lower clearance. The overall vertical dimension is
    $H=36$~mm, and the minimum jaw separation is
    $\Delta=9.2$~mm. The red and blue profiles represent
    the upper and lower jaw surfaces, respectively,
    while the dashed line indicates the $y=0$
    reference plane.}
    \label{fig:taper_offset}
\end{figure}

Wakefield simulations of collimators are often based on geometry files taken directly from the engineering design, where errors in the jaw position are not always apparent. Figure~\ref{fig:taper_offset} shows such a case for the ESR vertical collimator. In the initial input model the whole jaw assembly was displaced vertically with respect to the standard rectangular beam pipe, while the minimum jaw separation $\Delta = 2b$ and the overall height $H$ were preserved. As a result, the upper and lower halves of the collimator have different effective half-apertures $d \pm \delta$ and $b \pm \delta$, and the junctions between the displaced chamber and the beam pipe form untapered steps of height $\delta$. While the tapered jaws suppress the loss factor through the factor $[(2/\pi)\tan^{-1}(\mu d^2/\sigma_s L)]^2$, these steps are not tapered and contribute the full step loss. For long bunches they therefore dominate, and the simulated loss factor exceeds the prediction of Eq.~\eqref{eq:complete_model} by a large factor, far beyond the model uncertainty. The comparison with Eq.~\eqref{eq:complete_model} thus immediately identifies the inconsistency in the input model; after the jaw assembly was restored to its symmetric position, the simulated loss factor agreed with Eq.~\eqref{eq:complete_model}.

\section{Application to the EIC ESR Collimator}

The collimators of the EIC Electron Storage Ring (ESR) are designed with movable jaws. At high beam energy the jaws are going to be set closer to the beam, so the choice of material for the jaw tip becomes an important part of the design~\cite{Natochii2025,Natochii2026}. Here we consider only the geometric loss factor, due to the change in cross-section, assuming perfectly conducting walls; the resistive-wall contribution of the jaw-tip material is not included.

\begin{figure}[H]
\centering
\includegraphics[width=\columnwidth]{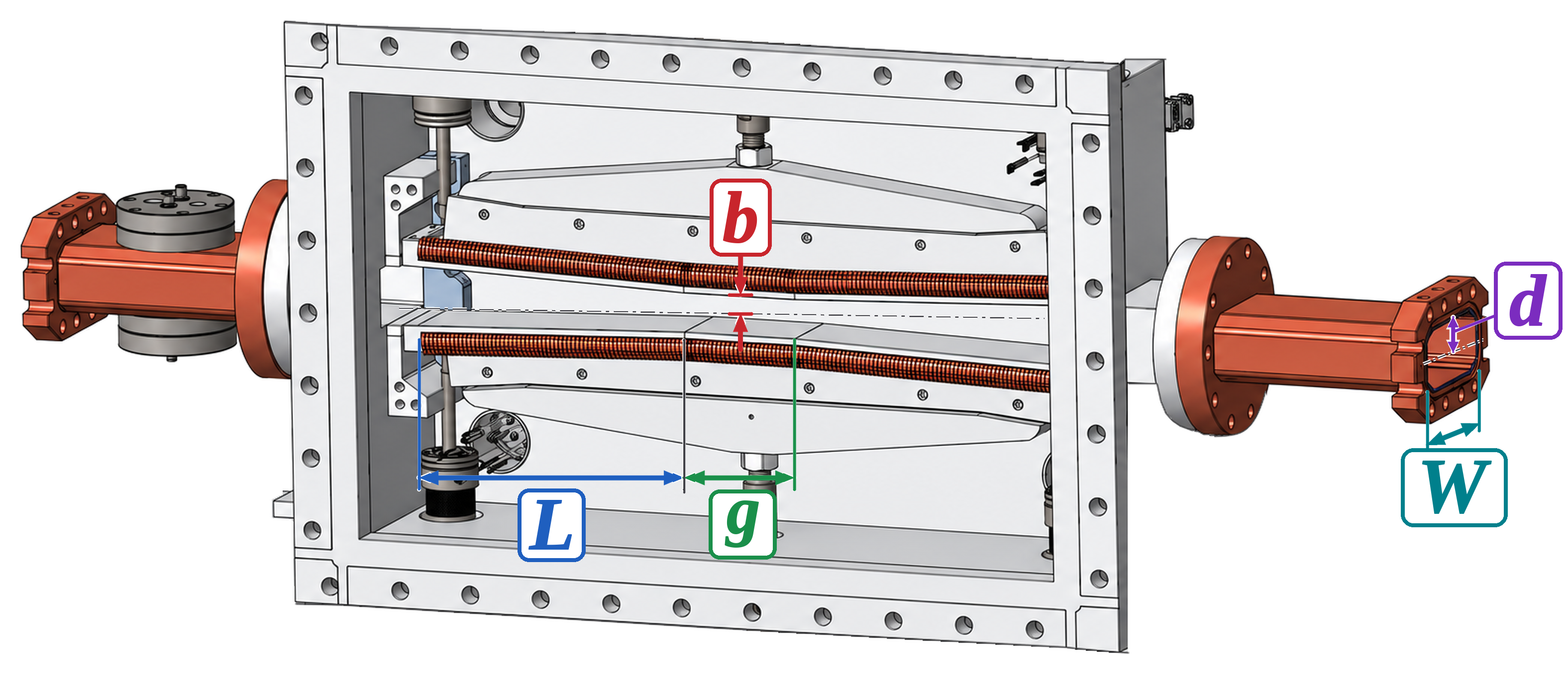}
\caption{Three-dimensional model of the EIC ESR flat
collimator, showing the opposing tapered jaws, vacuum chamber,
and beam-pipe connections.}
\label{fig:ESR_collimator}
\end{figure}

Figure~\ref{fig:ESR_collimator} shows the preliminary design of the vertical collimator. The geometry is characterized by $W=67$~mm, $d=18$~mm, $b=4.6$~mm, $L=300$~mm and $g=100$~mm, which are the chamber width, the maximum and minimum vertical half-apertures, the taper length and the length of the narrow straight section, respectively. Two vertical collimators with minimum vertical half-apertures of 4.6~mm and 6.5~mm are planned for operation at 9~GeV.

We use Eq.~\eqref{eq:complete_model} to determine the loss factor, $k_{\parallel,\mathrm{tpr}}$ of the flat tapered collimator. For the nominal rms bunch length $\sigma_s=7$~mm we have $N_F=d^2/(\sigma_s L)\approx0.15$ and $\sigma_s<\sigma_i=8.6$~mm, so the collimator operates in the optical regime, where the loss factor depends on $d^2/(\sigma_s L)$ and not on $b$. The minimum half-aperture enters only through the step loss factor $k_{\parallel,\mathrm{step}}(\alpha)$, $\alpha=b/d$. As a result, the loss factor changes only moderately with the jaw setting: $k_{\parallel,\mathrm{tpr}}=2.5$~mV/pC for $b=4.6$~mm and $1.9$~mV/pC for $b=6.5$~mm.

At 9~GeV, the ESR stores an average beam current of $I_\mathrm{avg}=2.5$~A in $M=1160$ bunches, with a bunch charge of $q_b=28$~nC, and a revolution period of $T_0=12.8$~$\mu$s. Assuming negligible multibunch wakefield interactions~\cite{Palumbo1995}, Eq.~\eqref{eq:complete_model} yields power losses~\cite{Chao1993} of $P_\mathrm{loss}
={k_{\parallel,\mathrm{tpr}} I_\mathrm{avg}^2 T_0}/{M}= 173$~W and $129$~W for $b=4.6$~mm and $b=6.5$~mm, respectively, at $\sigma_s=7$~mm. Resistive-wall losses have been analyzed separately
in Ref.~\cite{Natochii2026}.

\vspace{0.2cm}
\section{CONCLUSION}

In our earlier studies, we obtained a phenomenological expression for the loss factor of axisymmetric tapered collimators from numerical simulations~\cite{Blednykh2010} and subsequently extended it to rectangular geometries~\cite{Blednykh2011}. Two unexpected features were the dependence of the arctangent scaling parameter $\mu d^2/(\sigma_s L)$ on the larger aperture $d$, rather than the minimum aperture $b$, and the factor of two between the coefficients for axisymmetric and rectangular geometries. In the present work, we have provided a physical explanation for these observations. The Fresnel number interpretation clarifies the geometrical scaling, while the combination of diffraction theory and the electrodynamics of the beam-induced image charge explains the difference between the two geometries.

An important result of this work is the establishment of a loss-factor description for rectangular tapered collimators of finite chamber width and the extension of the phenomenological formulation from the optical to the inductive regime for both axisymmetric and rectangular geometries. In the optical limit, the loss factor of an abrupt step is described by the analytical expression derived by Bane, Stupakov and Zagorodnov~\cite{Bane2007}, whereas the reduction associated with tapering is represented by the phenomenological arctangent function. The extension to longer bunches is obtained using an analytical approximation based on the cutoff wavenumber of the lowest relevant electromagnetic mode. Interestingly, the larger aperture $d$ determines the characteristic scale in both the Fresnel-number description and the inductive approximation.

The resulting semi-analytical expression (Eq.~\eqref{eq:complete_model}) combines the analytical descriptions of the optical and inductive limits with the phenomenological reduction factor for a tapered transition. It provides a continuous description of the loss factor over a broad range of bunch lengths and geometrical parameters, in good agreement with ECHO and GdfidL simulations. Although the present analysis explains the geometrical scaling and the difference between axisymmetric and rectangular collimators, the specific arctangent dependence remains a numerical observation. An analytical derivation of this functional form from Maxwell's equations remains an open problem and would provide a more complete theoretical understanding of the loss factor of tapered structures.

\section*{ACKNOWLEDGMENTS}

This work was supported by Brookhaven Science Associates, LLC under Contract No. DE-SC0012704 with the U.S. Department of Energy. We are grateful to Michael Blaskiewicz for insightful discussions and guidance throughout this work. We thank Igor Zagorodnov for making the ECHO code available for our use. We also thank Warner Bruns for his valuable advice and assistance with GdfidL.


\bibliographystyle{apsrev4-2}
\bibliography{references}

\begin{thebibliography}{30}%
\makeatletter
\providecommand \@ifxundefined [1]{%
 \@ifx{#1\undefined}
}%
\providecommand \@ifnum [1]{%
 \ifnum #1\expandafter \@firstoftwo
 \else \expandafter \@secondoftwo
 \fi
}%
\providecommand \@ifx [1]{%
 \ifx #1\expandafter \@firstoftwo
 \else \expandafter \@secondoftwo
 \fi
}%
\providecommand \natexlab [1]{#1}%
\providecommand \enquote  [1]{``#1''}%
\providecommand \bibnamefont  [1]{#1}%
\providecommand \bibfnamefont [1]{#1}%
\providecommand \citenamefont [1]{#1}%
\providecommand \href@noop [0]{\@secondoftwo}%
\providecommand \href [0]{\begingroup \@sanitize@url \@href}%
\providecommand \@href[1]{\@@startlink{#1}\@@href}%
\providecommand \@@href[1]{\endgroup#1\@@endlink}%
\providecommand \@sanitize@url [0]{\catcode `\\12\catcode `\$12\catcode `\&12\catcode `\#12\catcode `\^12\catcode `\_12\catcode `\%12\relax}%
\providecommand \@@startlink[1]{}%
\providecommand \@@endlink[0]{}%
\providecommand \url  [0]{\begingroup\@sanitize@url \@url }%
\providecommand \@url [1]{\endgroup\@href {#1}{\urlprefix }}%
\providecommand \urlprefix  [0]{URL }%
\providecommand \Eprint [0]{\href }%
\providecommand \doibase [0]{https://doi.org/}%
\providecommand \selectlanguage [0]{\@gobble}%
\providecommand \bibinfo  [0]{\@secondoftwo}%
\providecommand \bibfield  [0]{\@secondoftwo}%
\providecommand \translation [1]{[#1]}%
\providecommand \BibitemOpen [0]{}%
\providecommand \bibitemStop [0]{}%
\providecommand \bibitemNoStop [0]{.\EOS\space}%
\providecommand \EOS [0]{\spacefactor3000\relax}%
\providecommand \BibitemShut  [1]{\csname bibitem#1\endcsname}%
\let\auto@bib@innerbib\@empty
\bibitem [{\citenamefont {Willeke}\ \emph {et~al.}(2021)\citenamefont {Willeke} \emph {et~al.}}]{EICCDR2021}%
  \BibitemOpen
  \bibfield  {author} {\bibinfo {author} {\bibfnamefont {F.}~\bibnamefont {Willeke}} \emph {et~al.},\ }\href {https://doi.org/10.2172/1765663} {\emph {\bibinfo {title} {Electron Ion Collider Conceptual Design Report 2021}}},\ \bibinfo {type} {Tech. Rep.}\ \bibinfo {number} {BNL-221006-2021-FORE}\ (\bibinfo  {institution} {Brookhaven National Laboratory},\ \bibinfo {address} {Upton, NY},\ \bibinfo {year} {2021})\BibitemShut {NoStop}%
\bibitem [{\citenamefont {Blednykh}\ and\ \citenamefont {Krinsky}(2010)}]{Blednykh2010}%
  \BibitemOpen
  \bibfield  {author} {\bibinfo {author} {\bibfnamefont {A.}~\bibnamefont {Blednykh}}\ and\ \bibinfo {author} {\bibfnamefont {S.}~\bibnamefont {Krinsky}},\ }\href {https://doi.org/10.1103/PhysRevSTAB.13.064401} {\bibfield  {journal} {\bibinfo  {journal} {Physical Review Special Topics - Accelerators and Beams}\ }\textbf {\bibinfo {volume} {13}},\ \bibinfo {pages} {064401} (\bibinfo {year} {2010})}\BibitemShut {NoStop}%
\bibitem [{\citenamefont {Blednykh}\ and\ \citenamefont {Krinsky}(2011)}]{Blednykh2011}%
  \BibitemOpen
  \bibfield  {author} {\bibinfo {author} {\bibfnamefont {A.}~\bibnamefont {Blednykh}}\ and\ \bibinfo {author} {\bibfnamefont {S.}~\bibnamefont {Krinsky}},\ }in\ \href@noop {} {\emph {\bibinfo {booktitle} {Proceedings of the 2011 Particle Accelerator Conference (PAC'11)}}}\ (\bibinfo {address} {New York, NY, USA},\ \bibinfo {year} {2011})\ pp.\ \bibinfo {pages} {1816--1818}\BibitemShut {NoStop}%
\bibitem [{\citenamefont {Blednykh}\ \emph {et~al.}(2021)\citenamefont {Blednykh}, \citenamefont {Bassi}, \citenamefont {Smaluk},\ and\ \citenamefont {Lindberg}}]{Blednykh2021}%
  \BibitemOpen
  \bibfield  {author} {\bibinfo {author} {\bibfnamefont {A.}~\bibnamefont {Blednykh}}, \bibinfo {author} {\bibfnamefont {G.}~\bibnamefont {Bassi}}, \bibinfo {author} {\bibfnamefont {V.}~\bibnamefont {Smaluk}},\ and\ \bibinfo {author} {\bibfnamefont {R.}~\bibnamefont {Lindberg}},\ }\href {https://doi.org/10.1103/PhysRevAccelBeams.24.104801} {\bibfield  {journal} {\bibinfo  {journal} {Physical Review Accelerators and Beams}\ }\textbf {\bibinfo {volume} {24}},\ \bibinfo {pages} {104801} (\bibinfo {year} {2021})}\BibitemShut {NoStop}%
\bibitem [{\citenamefont {Stupakov}\ \emph {et~al.}(2007)\citenamefont {Stupakov}, \citenamefont {Bane},\ and\ \citenamefont {Zagorodnov}}]{Stupakov2007Optical}%
  \BibitemOpen
  \bibfield  {author} {\bibinfo {author} {\bibfnamefont {G.}~\bibnamefont {Stupakov}}, \bibinfo {author} {\bibfnamefont {K.~L.~F.}\ \bibnamefont {Bane}},\ and\ \bibinfo {author} {\bibfnamefont {I.}~\bibnamefont {Zagorodnov}},\ }\href {https://doi.org/10.1103/PhysRevSTAB.10.054401} {\bibfield  {journal} {\bibinfo  {journal} {Physical Review Special Topics - Accelerators and Beams}\ }\textbf {\bibinfo {volume} {10}},\ \bibinfo {pages} {054401} (\bibinfo {year} {2007})}\BibitemShut {NoStop}%
\bibitem [{\citenamefont {Palumbo}\ \emph {et~al.}(1995)\citenamefont {Palumbo}, \citenamefont {Vaccaro},\ and\ \citenamefont {Zobov}}]{Palumbo1995}%
  \BibitemOpen
  \bibfield  {author} {\bibinfo {author} {\bibfnamefont {L.}~\bibnamefont {Palumbo}}, \bibinfo {author} {\bibfnamefont {V.~G.}\ \bibnamefont {Vaccaro}},\ and\ \bibinfo {author} {\bibfnamefont {M.}~\bibnamefont {Zobov}},\ }in\ \href@noop {} {\emph {\bibinfo {booktitle} {CAS -- CERN Accelerator School: Advanced Accelerator Physics}}},\ \bibinfo {series and number} {CERN 95-06}\ (\bibinfo {address} {Geneva},\ \bibinfo {year} {1995})\ pp.\ \bibinfo {pages} {331--390}\BibitemShut {NoStop}%
\bibitem [{\citenamefont {Chao}(1993)}]{Chao1993}%
  \BibitemOpen
  \bibfield  {author} {\bibinfo {author} {\bibfnamefont {A.~W.}\ \bibnamefont {Chao}},\ }\href@noop {} {\emph {\bibinfo {title} {Physics of Collective Beam Instabilities in High Energy Accelerators}}}\ (\bibinfo  {publisher} {John Wiley \& Sons},\ \bibinfo {address} {New York},\ \bibinfo {year} {1993})\BibitemShut {NoStop}%
\bibitem [{\citenamefont {Zotter}\ and\ \citenamefont {Kheifets}(1998)}]{Zotter1998}%
  \BibitemOpen
  \bibfield  {author} {\bibinfo {author} {\bibfnamefont {B.~W.}\ \bibnamefont {Zotter}}\ and\ \bibinfo {author} {\bibfnamefont {S.~A.}\ \bibnamefont {Kheifets}},\ }\href@noop {} {\emph {\bibinfo {title} {Impedances and Wakes in High-Energy Particle Accelerators}}}\ (\bibinfo  {publisher} {World Scientific},\ \bibinfo {year} {1998})\BibitemShut {NoStop}%
\bibitem [{\citenamefont {Yokoya}(1990)}]{Yokoya1990}%
  \BibitemOpen
  \bibfield  {author} {\bibinfo {author} {\bibfnamefont {K.}~\bibnamefont {Yokoya}},\ }\href@noop {} {\emph {\bibinfo {title} {Impedance of Slowly Tapered Structures}}},\ \bibinfo {type} {Tech. Rep.}\ \bibinfo {number} {CERN-SL-90-88-AP}\ (\bibinfo  {institution} {CERN},\ \bibinfo {year} {1990})\BibitemShut {NoStop}%
\bibitem [{\citenamefont {Stupakov}(2007)}]{Stupakov2007}%
  \BibitemOpen
  \bibfield  {author} {\bibinfo {author} {\bibfnamefont {G.}~\bibnamefont {Stupakov}},\ }\href {https://doi.org/10.1103/PhysRevSTAB.10.094401} {\bibfield  {journal} {\bibinfo  {journal} {Phys. Rev. ST Accel. Beams}\ }\textbf {\bibinfo {volume} {10}},\ \bibinfo {pages} {094401} (\bibinfo {year} {2007})}\BibitemShut {NoStop}%
\bibitem [{\citenamefont {Bane}\ \emph {et~al.}(2007)\citenamefont {Bane}, \citenamefont {Stupakov},\ and\ \citenamefont {Zagorodnov}}]{Bane2007}%
  \BibitemOpen
  \bibfield  {author} {\bibinfo {author} {\bibfnamefont {K.~L.~F.}\ \bibnamefont {Bane}}, \bibinfo {author} {\bibfnamefont {G.}~\bibnamefont {Stupakov}},\ and\ \bibinfo {author} {\bibfnamefont {I.}~\bibnamefont {Zagorodnov}},\ }\href {https://doi.org/10.1103/PhysRevSTAB.10.074401} {\bibfield  {journal} {\bibinfo  {journal} {Physical Review Special Topics - Accelerators and Beams}\ }\textbf {\bibinfo {volume} {10}},\ \bibinfo {pages} {074401} (\bibinfo {year} {2007})}\BibitemShut {NoStop}%
\bibitem [{\citenamefont {Stupakov}\ and\ \citenamefont {Podobedov}(2010)}]{StupakovPodobedov2010}%
  \BibitemOpen
  \bibfield  {author} {\bibinfo {author} {\bibfnamefont {G.}~\bibnamefont {Stupakov}}\ and\ \bibinfo {author} {\bibfnamefont {B.}~\bibnamefont {Podobedov}},\ }\href {https://doi.org/10.1103/PhysRevSTAB.13.104401} {\bibfield  {journal} {\bibinfo  {journal} {Phys. Rev. ST Accel. Beams}\ }\textbf {\bibinfo {volume} {13}},\ \bibinfo {pages} {104401} (\bibinfo {year} {2010})}\BibitemShut {NoStop}%
\bibitem [{\citenamefont {Zagorodnov}\ and\ \citenamefont {Weiland}(2005)}]{Zagorodnov2005}%
  \BibitemOpen
  \bibfield  {author} {\bibinfo {author} {\bibfnamefont {I.}~\bibnamefont {Zagorodnov}}\ and\ \bibinfo {author} {\bibfnamefont {T.}~\bibnamefont {Weiland}},\ }\href {https://doi.org/10.1103/PhysRevSTAB.8.042001} {\bibfield  {journal} {\bibinfo  {journal} {Phys. Rev. ST Accel. Beams}\ }\textbf {\bibinfo {volume} {8}},\ \bibinfo {pages} {042001} (\bibinfo {year} {2005})}\BibitemShut {NoStop}%
\bibitem [{\citenamefont {Zagorodnov}(2006)}]{Zagorodnov2006}%
  \BibitemOpen
  \bibfield  {author} {\bibinfo {author} {\bibfnamefont {I.}~\bibnamefont {Zagorodnov}},\ }\href {https://doi.org/10.1103/PhysRevSTAB.9.102002} {\bibfield  {journal} {\bibinfo  {journal} {Phys. Rev. ST Accel. Beams}\ }\textbf {\bibinfo {volume} {9}},\ \bibinfo {pages} {102002} (\bibinfo {year} {2006})}\BibitemShut {NoStop}%
\bibitem [{\citenamefont {Bruns}(nd)}]{BrunsGdfidL}%
  \BibitemOpen
  \bibfield  {author} {\bibinfo {author} {\bibfnamefont {W.}~\bibnamefont {Bruns}},\ }\href@noop {} {\bibinfo {title} {{GdfidL}: Electromagnetic field simulator}},\ \bibinfo {howpublished} {\url{https://www.gdfidl.de/}} (\bibinfo {year} {n.d.}),\ \bibinfo {note} {accessed: 2026-09-27}\BibitemShut {NoStop}%
\bibitem [{\citenamefont {Siegman}(1986)}]{Siegman1986}%
  \BibitemOpen
  \bibfield  {author} {\bibinfo {author} {\bibfnamefont {A.~E.}\ \bibnamefont {Siegman}},\ }\href@noop {} {\emph {\bibinfo {title} {Lasers}}}\ (\bibinfo  {publisher} {University Science Books},\ \bibinfo {address} {Mill Valley, CA},\ \bibinfo {year} {1986})\BibitemShut {NoStop}%
\bibitem [{\citenamefont {Zhao}\ \emph {et~al.}(2023)\citenamefont {Zhao}, \citenamefont {Tao}, \citenamefont {Weber}, \citenamefont {Kaune}, \citenamefont {Schuster}, \citenamefont {Hao},\ and\ \citenamefont {Wanner}}]{FresnelDiffraction}%
  \BibitemOpen
  \bibfield  {author} {\bibinfo {author} {\bibfnamefont {M.}~\bibnamefont {Zhao}}, \bibinfo {author} {\bibfnamefont {Y.}~\bibnamefont {Tao}}, \bibinfo {author} {\bibfnamefont {K.}~\bibnamefont {Weber}}, \bibinfo {author} {\bibfnamefont {T.}~\bibnamefont {Kaune}}, \bibinfo {author} {\bibfnamefont {S.}~\bibnamefont {Schuster}}, \bibinfo {author} {\bibfnamefont {Z.}~\bibnamefont {Hao}},\ and\ \bibinfo {author} {\bibfnamefont {G.}~\bibnamefont {Wanner}},\ }\href {https://doi.org/10.3390/s23229024} {\bibfield  {journal} {\bibinfo  {journal} {Sensors}\ }\textbf {\bibinfo {volume} {23}},\ \bibinfo {pages} {9024} (\bibinfo {year} {2023})}\BibitemShut {NoStop}%
\bibitem [{\citenamefont {Paschotta}(nd)}]{PaschottaFresnel}%
  \BibitemOpen
  \bibfield  {author} {\bibinfo {author} {\bibfnamefont {R.}~\bibnamefont {Paschotta}},\ }\href {https://doi.org/10.61835/uvi} {\bibinfo {title} {Fresnel number}},\ \bibinfo {howpublished} {RP Photonics Encyclopedia} (\bibinfo {year} {n.d.}),\ \bibinfo {note} {accessed October 23, 2026}\BibitemShut {NoStop}%
\bibitem [{\citenamefont {Lin}\ and\ \citenamefont {Whittum}(2000)}]{Lin2000}%
  \BibitemOpen
  \bibfield  {author} {\bibinfo {author} {\bibfnamefont {X.~E.}\ \bibnamefont {Lin}}\ and\ \bibinfo {author} {\bibfnamefont {D.~H.}\ \bibnamefont {Whittum}},\ }\href {https://doi.org/10.1103/PhysRevSTAB.3.101001} {\bibfield  {journal} {\bibinfo  {journal} {Phys. Rev. ST Accel. Beams}\ }\textbf {\bibinfo {volume} {3}},\ \bibinfo {pages} {101001} (\bibinfo {year} {2000})}\BibitemShut {NoStop}%
\bibitem [{\citenamefont {Balakin}\ and\ \citenamefont {Novokhatski}(1983)}]{Balakin1983}%
  \BibitemOpen
  \bibfield  {author} {\bibinfo {author} {\bibfnamefont {V.~E.}\ \bibnamefont {Balakin}}\ and\ \bibinfo {author} {\bibfnamefont {A.~V.}\ \bibnamefont {Novokhatski}},\ }in\ \href@noop {} {\emph {\bibinfo {booktitle} {Proceedings of the 12th International Conference on High-Energy Accelerators}}},\ \bibinfo {editor} {edited by\ \bibinfo {editor} {\bibfnamefont {F.~T.}\ \bibnamefont {Cole}}\ and\ \bibinfo {editor} {\bibfnamefont {R.}~\bibnamefont {Donaldson}}}\ (\bibinfo  {publisher} {Fermilab},\ \bibinfo {address} {Batavia, IL},\ \bibinfo {year} {1983})\ p.\ \bibinfo {pages} {117}\BibitemShut {NoStop}%
\bibitem [{\citenamefont {Bisognano}\ \emph {et~al.}(1988)\citenamefont {Bisognano}, \citenamefont {Heifets},\ and\ \citenamefont {Yunn}}]{Bisognano1988}%
  \BibitemOpen
  \bibfield  {author} {\bibinfo {author} {\bibfnamefont {J.~J.}\ \bibnamefont {Bisognano}}, \bibinfo {author} {\bibfnamefont {S.~A.}\ \bibnamefont {Heifets}},\ and\ \bibinfo {author} {\bibfnamefont {B.~C.}\ \bibnamefont {Yunn}},\ }in\ \href@noop {} {\emph {\bibinfo {booktitle} {Proceedings of the European Particle Accelerator Conference (EPAC 1988)}}}\ (\bibinfo {address} {Rome, Italy},\ \bibinfo {year} {1988})\ pp.\ \bibinfo {pages} {708--710}\BibitemShut {NoStop}%
\bibitem [{\citenamefont {Heifets}\ and\ \citenamefont {Kheifets}(1991)}]{Heifets1991}%
  \BibitemOpen
  \bibfield  {author} {\bibinfo {author} {\bibfnamefont {S.~A.}\ \bibnamefont {Heifets}}\ and\ \bibinfo {author} {\bibfnamefont {S.~A.}\ \bibnamefont {Kheifets}},\ }\href {https://doi.org/10.1103/RevModPhys.63.631} {\bibfield  {journal} {\bibinfo  {journal} {Rev. Mod. Phys.}\ }\textbf {\bibinfo {volume} {63}},\ \bibinfo {pages} {631} (\bibinfo {year} {1991})}\BibitemShut {NoStop}%
\bibitem [{\citenamefont {Olver}\ \emph {et~al.}(2010)\citenamefont {Olver}, \citenamefont {Lozier}, \citenamefont {Boisvert},\ and\ \citenamefont {Clark}}]{Olver2010}%
  \BibitemOpen
  \bibinfo {editor} {\bibfnamefont {F.~W.~J.}\ \bibnamefont {Olver}}, \bibinfo {editor} {\bibfnamefont {D.~W.}\ \bibnamefont {Lozier}}, \bibinfo {editor} {\bibfnamefont {R.~F.}\ \bibnamefont {Boisvert}},\ and\ \bibinfo {editor} {\bibfnamefont {C.~W.}\ \bibnamefont {Clark}},\ eds.,\ \href@noop {} {\emph {\bibinfo {title} {NIST Handbook of Mathematical Functions}}}\ (\bibinfo  {publisher} {Cambridge University Press},\ \bibinfo {address} {Cambridge, UK},\ \bibinfo {year} {2010})\BibitemShut {NoStop}%
\bibitem [{{DLMF}()}]{DLMF}%
  \BibitemOpen
  {DLMF},\ \href@noop {} {\bibinfo {title} {{NIST Digital Library of Mathematical Functions}}},\ \bibinfo {howpublished} {\url{https://dlmf.nist.gov/}} (\bibinfo {year} {2026}),\ \bibinfo {note} {release 1.2.8 of 2026-09-15; F. W. J. Olver, A. B. Olde Daalhuis, D. W. Lozier, B. I. Schneider, R. F. Boisvert, C. W. Clark, B. R. Miller, B. V. Saunders, H. S. Cohl, and M. A. McClain, eds.}\BibitemShut {Stop}%
\bibitem [{\citenamefont {Blednykh}\ and\ \citenamefont {Krinsky}(2012)}]{Blednykh2012}%
  \BibitemOpen
  \bibfield  {author} {\bibinfo {author} {\bibfnamefont {A.}~\bibnamefont {Blednykh}}\ and\ \bibinfo {author} {\bibfnamefont {S.}~\bibnamefont {Krinsky}},\ }\href {https://doi.org/10.1103/PhysRevSTAB.15.054405} {\bibfield  {journal} {\bibinfo  {journal} {Physical Review Special Topics - Accelerators and Beams}\ }\textbf {\bibinfo {volume} {15}},\ \bibinfo {pages} {054405} (\bibinfo {year} {2012})}\BibitemShut {NoStop}%
\bibitem [{\citenamefont {Blednykh}\ and\ \citenamefont {Krinsky}(2013)}]{Blednykh2013}%
  \BibitemOpen
  \bibfield  {author} {\bibinfo {author} {\bibfnamefont {A.}~\bibnamefont {Blednykh}}\ and\ \bibinfo {author} {\bibfnamefont {S.}~\bibnamefont {Krinsky}},\ }in\ \href@noop {} {\emph {\bibinfo {booktitle} {Proceedings of PAC2013}}}\ (\bibinfo {address} {Pasadena, CA, USA},\ \bibinfo {year} {2013})\BibitemShut {NoStop}%
\bibitem [{\citenamefont {Stupakov}(2009)}]{Stupakov2009}%
  \BibitemOpen
  \bibfield  {author} {\bibinfo {author} {\bibfnamefont {G.}~\bibnamefont {Stupakov}},\ }in\ \href@noop {} {\emph {\bibinfo {booktitle} {Proceedings of PAC09}}}\ (\bibinfo {address} {Vancouver, B.C., Canada},\ \bibinfo {year} {2009})\BibitemShut {NoStop}%
\bibitem [{\citenamefont {Morton}(1993)}]{Morton1993}%
  \BibitemOpen
  \bibfield  {author} {\bibinfo {author} {\bibfnamefont {P.~L.}\ \bibnamefont {Morton}},\ }\href {https://www.slac.stanford.edu/pubs/slacpubs/6000/slac-pub-6052.pdf} {\emph {\bibinfo {title} {Introduction to Impedance for Short Relativistic Bunches}}},\ \bibinfo {type} {Tech. Rep.}\ \bibinfo {number} {SLAC-PUB-6052}\ (\bibinfo  {institution} {Stanford Linear Accelerator Center},\ \bibinfo {year} {1993})\BibitemShut {NoStop}%
\bibitem [{\citenamefont {Natochii}\ \emph {et~al.}(2025)\citenamefont {Natochii}, \citenamefont {Aschenauer}, \citenamefont {Hamdi}, \citenamefont {Hetzel}, \citenamefont {Llink}, \citenamefont {Marx}, \citenamefont {Montag}, \citenamefont {Tepikian}, \citenamefont {Cai},\ and\ \citenamefont {Nosochkov}}]{Natochii2025}%
  \BibitemOpen
  \bibfield  {author} {\bibinfo {author} {\bibfnamefont {A.}~\bibnamefont {Natochii}}, \bibinfo {author} {\bibfnamefont {E.-C.}\ \bibnamefont {Aschenauer}}, \bibinfo {author} {\bibfnamefont {K.}~\bibnamefont {Hamdi}}, \bibinfo {author} {\bibfnamefont {C.}~\bibnamefont {Hetzel}}, \bibinfo {author} {\bibfnamefont {E.}~\bibnamefont {Llink}}, \bibinfo {author} {\bibfnamefont {D.}~\bibnamefont {Marx}}, \bibinfo {author} {\bibfnamefont {C.}~\bibnamefont {Montag}}, \bibinfo {author} {\bibfnamefont {S.}~\bibnamefont {Tepikian}}, \bibinfo {author} {\bibfnamefont {Y.}~\bibnamefont {Cai}},\ and\ \bibinfo {author} {\bibfnamefont {Y.}~\bibnamefont {Nosochkov}},\ }\href@noop {} {\bibfield  {journal} {\bibinfo  {journal} {arXiv preprint}\ } (\bibinfo {year} {2025})},\ \Eprint {https://arxiv.org/abs/2512.19502} {arXiv:2512.19502} \BibitemShut {NoStop}%
\bibitem [{\citenamefont {Natochii}\ \emph {et~al.}(2026)\citenamefont {Natochii}, \citenamefont {Aschenauer}, \citenamefont {Best}, \citenamefont {Blednykh}, \citenamefont {Gu}, \citenamefont {Gunavathi}, \citenamefont {Hetzel},\ and\ \citenamefont {Montag}}]{Natochii2026}%
  \BibitemOpen
  \bibfield  {author} {\bibinfo {author} {\bibfnamefont {A.}~\bibnamefont {Natochii}}, \bibinfo {author} {\bibfnamefont {E.-C.}\ \bibnamefont {Aschenauer}}, \bibinfo {author} {\bibfnamefont {J.}~\bibnamefont {Best}}, \bibinfo {author} {\bibfnamefont {A.}~\bibnamefont {Blednykh}}, \bibinfo {author} {\bibfnamefont {X.}~\bibnamefont {Gu}}, \bibinfo {author} {\bibfnamefont {K.}~\bibnamefont {Gunavathi}}, \bibinfo {author} {\bibfnamefont {C.}~\bibnamefont {Hetzel}},\ and\ \bibinfo {author} {\bibfnamefont {C.}~\bibnamefont {Montag}},\ }\bibfield  {journal} {\bibinfo  {journal} {arXiv preprint}\ }\href {https://doi.org/10.48550/arXiv.2610.04682} {10.48550/arXiv.2610.04682} (\bibinfo {year} {2026}),\ \Eprint {https://arxiv.org/abs/2610.04682} {arXiv:2610.04682 [physics.acc-ph]} \BibitemShut {NoStop}%
\end{thebibliography}%

\end{document}